\documentclass[12pt,a4paper,nofootinbib]{revtex4}

\usepackage[T1]{fontenc}
\usepackage{lmodern}

\usepackage[T1]{fontenc}
\usepackage{lmodern}

\usepackage{amsmath}
\usepackage{amssymb}
\usepackage{mathtools}
\usepackage{graphicx}
\usepackage{relsize}
\usepackage{exscale}
\usepackage{wasysym}
\usepackage{ulem}
\usepackage{cancel}
\usepackage{titlesec}
\usepackage{float}
\usepackage{enumitem}
\usepackage{graphicx}
\usepackage{slashed}
\usepackage[left=2cm,right=2cm,top=2cm,bottom=2cm]{geometry}

\begin{document}

\rightline{UTP-UU-13/20, SPIN-13/14}

 \title{Backreaction of a massless minimally coupled scalar field
        from inflationary quantum fluctuations}

\author{Dra\v{z}en Glavan$^*$  
       }

\author{Tomislav Prokopec$^*$  
        }

\author{Vasileios Prymidis\footnote{E-mail: \tt d.glavan@uu.nl; 
                t.prokopec@uu.nl; v.prymidis@students.uu.nl}
       }

\affiliation{Institute for Theoretical Physics and Spinoza Institute, Utrecht University,\\
Postbus 80.195, 3508 TD Utrecht, The Netherlands}

%
%
\begin{abstract} 

In this paper we study a massless, minimally coupled scalar field in 
a FLRW spacetime with periods of different constant deceleration parameter. 
We assume the Bunch-Davies vacuum during inflation and then 
use a sudden matching approximation to match it
onto radiation era and subsequently onto matter era. 
We then proceed to calculate the one-loop energy-momentum tensor
from the inflationary quantum vacuum fluctuations in different eras. 
The energy-momentum tensor
has the form of an ideal (quantum) fluid, characterized by an equation
of state. When compared with the background, far away from the matching 
the quantum energy density in radiation era exhibits a contribution 
that grows logarithmically with the scale factor.
In matter era the ratio of the quantum to classical fluid settles eventually 
to a tiny constant, 
$\rho_q/\rho \simeq (\hbar H_0)^2/(4\pi m_{\rm P}^2 c^4)\sim 10^{-13}$ 
for a grand unified scale inflation.
Curiously, the late time scaling of quantum fluctuations suggests 
that they contribute a little to the dark matter of the Universe,
provided that it clusters as cold dark matter, which needs to be checked. 

\end{abstract}

\maketitle

\section{Introduction and motivation}

 There has been some excitement concerning the possibility that 
inflationary vacuum fluctuations can play the role of dark energy 
today~\cite{KolbMatarreseRiotto, BarausseMatarreseRiotto}.
 It was subsequently argued that that was not the case \cite{HirataSeljak}. 
That question was never completely settled by performing a proper 
quantum field theoretic calculation.

 Somewhat earlier, in the early 2000s, 
some progress has been made as regards the
(one-loop) backreaction of scalar cosmological perturbations
during inflation~\cite{Abramo:2001dc,Abramo:2001db,Abramo:1998hj,Abramo:1998hi,
Geshnizjani:2002wp,Geshnizjani:2003cn,Abramo:2001dd}.
The question of backreaction is particularly delicate during inflation since 
the main contribution to the backreaction during this era
originates from super-Hubble scales and different observers will
disagree regarding how to define the backreaction. 
A consensus has been reached that a reasonable (local) observer will 
observe a local expansion 
rate~\cite{Geshnizjani:2002wp,Geshnizjani:2003cn,Abramo:2001dd},
or an expansion rate that contains information collected along the past light
cone~\cite{Abramo:2001dc,Abramo:2001db,Abramo:1998hj,Abramo:1998hi}.
In both viewpoints 
the same conclusion was reached: at one-loop cosmological perturbations
produce {\it vanishing} backreaction. This conclusion was 
disappointing and has contributed to stopping any further investigation 
of quantum backreaction, including that on late time observers
\footnote{That a significant backgreaction in cosmological spaces might occur
was suggested by the studies of Janssen and 
Prokopec~\cite{Janssen:2009nz,Janssen:2008dw}, in which the authors considered 
the backreaction induced at one-loop order by non-minimally coupled scalars 
and in spaces with a constant deceleration parameter $q=\epsilon-1$, where 
$\epsilon$ denotes a slow roll parameter. Inspired by an erlier 
work~ of Ford and Vilenkin~\cite{Vilenkin:1982wt,Ford:1985qh}, 
the authors regularised the infrared by matching onto 
a preceding radiation era in which the Bunch-Davies is regular in the 
infrared and found that for certain $\epsilon$'s the relative one-loop
contribution to the energy density and pressure grows when compared 
with the background densities.}.

 What holds for inflationary observers is not generally true for late time
observers. Namely, at late times many of the modes that were super-Hubble
during inflation, become sub-Hubble, and they may have an impact 
on observations. The main purpose of this paper is to study the one-loop
quantum backreaction of amplified inflationary fluctuations in late time
epochs such as radiation and matter era. For simplicity, we concentrate here 
on the following, somewhat simplified case:
we assume a de Sitter inflation, suddenly matched onto a radiation era,
that is subsequently matched onto a matter era. 
We consider quantum fluctuations 
of a massless, minimally coupled test scalar field and of the graviton,
and calculate the corresponding one-loop energy-momentum tensor, which we 
use to study quantum backreaction. Furthermore, we assume a global 
Bunch-Davies vacuum state during inflation, and then evolve it 
by performing the usual sudden matching onto radiation and matter era.  

\begin{figure}[hh]
\centering
\includegraphics[scale=0.68]{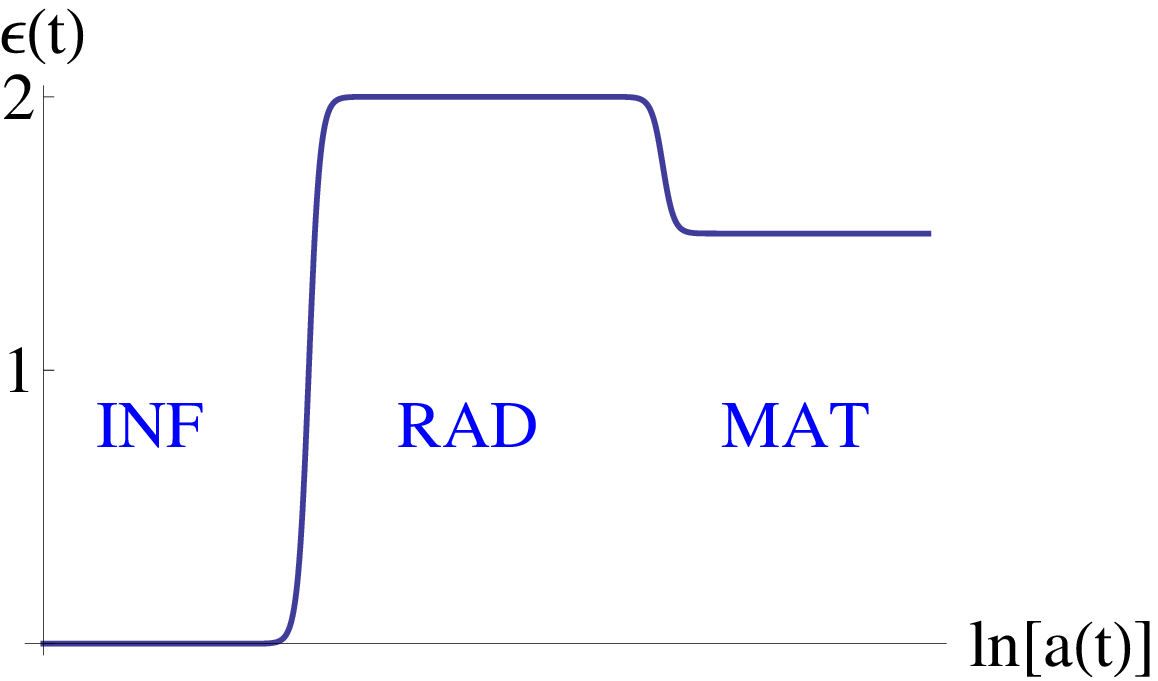}
\hskip 0.5cm
\includegraphics[scale=0.74]{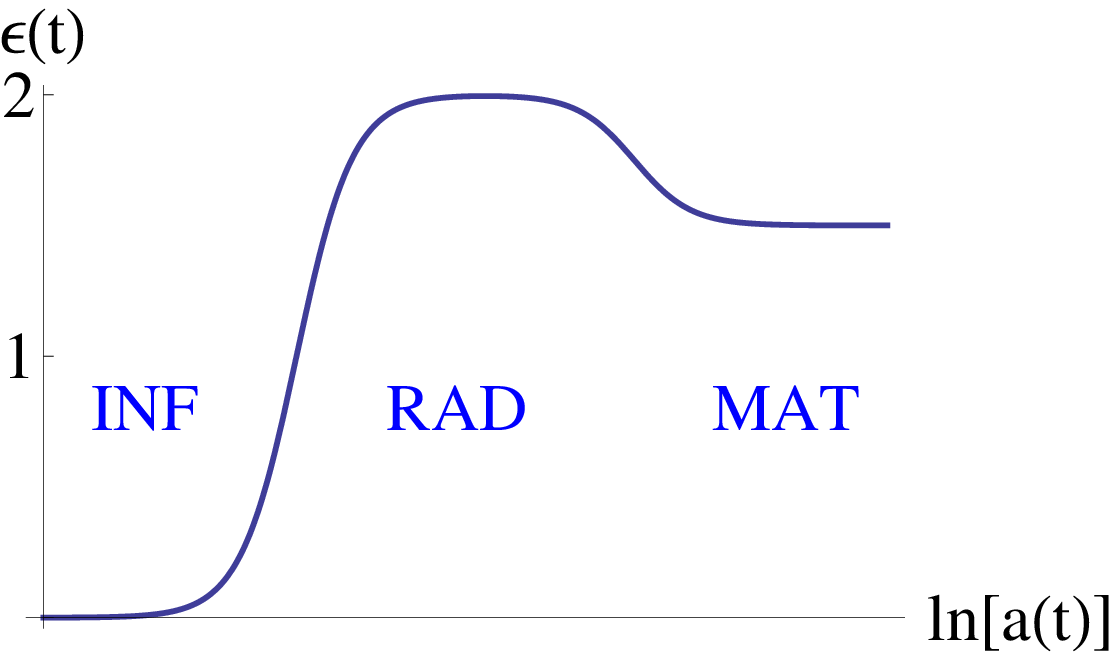}
\caption{Two possible (realistic) scenarios for the evolution of 
the slow roll parameter $\epsilon(t)$ in de Sitter inflation, 
radiation and matter era. 
{\it Left panel:}  Sharp transitions, in which the transition time $\tau$
is smaller than the expansion time, $\tau\ll 1/H$.
{\it Right panel:}  Mild transitions, in which $\tau \gtrsim 1/H$.
}
\label{figure 1:epsilon}
\end{figure}

With some care we point out the difficulties that arise from this 
procedure. Firstly, strictly speaking the global Bunch-Davies vacuum
state we use is a singular state in the infrared (IR), 
making it thus unphysical. For our purposes this state suffices, 
since the singular nature of the state 
is not apparent in the one-loop energy-momentum tensor.
But the singular nature of the global Bunch-Davies state 
can become apparent even in a one-loop energy-momentum tensor.
For example, the global Bunch-Davies vacuum in matter era 
exhibits a logarithmic infrared divergence in the one-loop 
energy-momentum, 
and the state has to be changed (`regularized') in the infrared. 
Secondly, we perform a sudden matching in between inflation, radiation and 
matter eras (see figure~\ref{figure 1:epsilon}) 
in the sense that the background geometry exhibits a jump 
(components of the curvature tensors exhibit a jump).
This {\it singular} procedure generates extra ultraviolet (UV) divergences, which are 
manifest in dimensional regularization as an extra 
divergence $\propto 1/(D-4)$, that scales as radiation,
and thus cannot be renormalized by the usual (bulk) counterterms.
Furthermore, there are power law boundary divergences
that have the same origin, and that are regular at late times. 
All of these divergences are unphysical, in the sense that they are 
generated by the unphysical procedure of sudden matching. 
An example which demonstrates by an exemplary calculation that 
a smooth transition does not generate any boundary divergences is discussed
in Ref.~\cite{Koivisto:2010pj}, but we do not see how that exact solution
can be extended to the realistic Universe history we are interested
to model in this work. Instead, we 
introduce an exponential regulator for the Bogolyubov coefficients, 
which mimics a finite time of transition, thus modeling  
the physically realistic situation of a smooth transition,
and thus demonstrate that all of these boundary divergences are removed.  
The transition time in the Universe is typically given by 
the expansion time at that moment. 
For example, the radiation-matter transition 
time is of the order of the expansion time at the time of radiation-matter 
equality. The analogous estimate for the inflation-radiation transition is 
also reasonable: 
the time scale of the inflaton decay at the end of inflation 
(known as (p)reheating) is the expansion time at the end of inflation.
However, when non-perturbative effects such as parametric resonance are 
taken into account, the transition time can become of the 
order tens of oscillations of the inflaton (dictated by the inflaton mass), 
which can be much shorter than the expansion time at that time.
Hence, since the (p)reheating time is not dictated by the expansion rate, 
(p)reheating can last much shorter (but also much longer)
than the expansion time at the end of inflation, and should be kept as 
a free parameter (see, for example, Ref.~\cite{Prokopec:1996rr}). 

 In this work we calculate the expectation value of the one-loop
energy-momentum tensor with respect to a globally defined Bunch-Davies vacuum,
and hence any effect that we calculate on the background geometry is of 
a global character. However, observations are done in our local Universe,
and are limited by causality to be within our Hubble volume. The question
then arises whether these global effects can have any effect on our local
measurements. The answer is affirmative. Namely, 
a photon moving in a global space-time governed by a global expansion rate 
$H(t)$ will exhibit a redshift that corresponds to that expansion rate.
Of course, on top of this global effect, the photon will also feel 
the presence of any local fluctuations around $H(t)$.
Let us for the sake of argument refer to these local fluctuations as
$\delta H(x)$ (a better way of parametrizing this is by making use of the local
expansion parameter from the Raychaudhuri equation, 
see \cite{Geshnizjani:2002wp,Geshnizjani:2003cn}).
In this work we do not attempt to estimate the size of these
local fluctuations in the expansion rate. However, the procedure of 
how to estimate their size is clear: one has to calculate 
the one-loop energy-momentum -- energy-momentum  correlator, from which 
one would be able to extract the energy density correlator, 
$\langle \delta\rho(x)\delta\rho(x^{\prime})/\rho_0^2(t)\rangle$.
This correlator then contains enough information to extract how the  
Hubble rate fluctuates around the global Hubble rate calculated in this paper. 
The statement that, on average, photons will exhibit 
a net change in the expansion rate, and hence a net redshift, 
stands firm.

 The paper is organized as follows. In section~\ref{FLRW background}
we review the background geometry of a spatially flat universe.
This is followed in ~\ref{Scalar field on FLRW} by a brief account of 
the scalar field quantization on a homogeneous expanding background.
The one-loop expression for the scalar energy-momentum tensor
is presented in section~\ref{Energy-momentum tensor}.
In section~\ref{Energy-momentum tensor on FLRW with constant epsilon}
we present the results for the one-loop energy-momentum tensor 
in de Sitter, radiation and matter era for a global Bunch-Davies 
vacuum (some details of this standard calculation are given in Appendix~A). 
Our main results can be found in 
section~\ref{Energy-momentum on FLRW with evolving epsilon},
where we discuss the one-loop energy-momentum tensor, and the resulting 
backreaction on the background geometry, in radiation and matter era. 
A particular attention is devoted to resolving the effects arising from
the sudden matching approximation. Finally, 
in section~\ref{Discussion and conclusions} we conclude.

\section{FLRW background}
\label{FLRW background}

A spatially flat Friedmann-Lema\^{i}tre-Robertson-Walker (FLRW) metric, 
$g_{\mu\nu}$, in $D$ dimensions, written in conformal time, is
\begin{equation}
g_{\mu\nu}(x) = a^2(\eta) \eta_{\mu\nu} 
\, ,
\end{equation}
where $\eta_{\mu\nu}$ is the Minkowski metric in $D$ dimensions,
\begin{equation}
\eta_{\mu\nu} = \mathrm{diag}(-1,1,1,\dots,1) 
\, ,
\end{equation}
$a(\eta)$ is the scale factor describing the time evolution of physical
length scales, and conformal time $\eta$ is related to the physical
(cosmological) time $t$ via $dt=a(\eta)d\eta$. 
The reason we are working in conformal time, instead of physical one,
is that we find the calculations simpler to perform. 
We introduce the conformal Hubble rate
\begin{equation}
\mathcal{H}(\eta) = \frac{a'(\eta)}{a(\eta)} 
\, ,
\end{equation}
which is related to the usual physical Hubble rate by 
$\mathcal{H}=aH$, $H=\dot{a}/a$, where a {\it prime} denotes a differentiation 
with respect to conformal time, and a {\it dot} a differentiation with respect
to physical time. The dynamics of this background are described by
the Friedmann equations, 
\begin{align}
 & \left( \frac{\mathcal{H}}{a} \right)^2 
      = \frac{8\pi G_N}{3c^2} \rho + \frac{\Lambda}{3} \ , \\
& \frac{\mathcal{H}'-\mathcal{H}^2}{a^2} = -\frac{4\pi G_N}{c^2} (\rho+p)
\, ,
\end{align}
where $\rho$ and $p$ are the energy density and pressure of the background 
ideal fluid, respectively, $\Lambda$ is the cosmological constant, 
$G_N$ is the Newton's constant, and $c$ the speed of light.
We are not interested in the physical composition of the classical fluid
that causes the Universe to expand.
We are only concerned with what is its effective equation of state, {\it i.e.}
how the equation of state parameter $w=p/\rho$ depends on time. 
Furthermore, we are not interested in fluctuations of the fluid; namely we 
shall assume that the classical fluid is perfectly homogeneous
(that defines the hyper-surface of simultaneity), and that it 
does not fluctuate. In our toy model, the sole source of (quantum) fluctuations
is generated by a massless scalar field.
Namely, we shall canonically quantize the field on this fixed background,
such that quantum fluctuations of the scalar field 
are the sole source of density and pressure fluctuations. In addition, 
we shall present a quantitative argument on how to include 
the contribution of one-loop (quantum) graviton fluctuations. 

From the covariant conservation law, 
\begin{equation}
\rho^\prime + 3{\cal H}(\rho+p)=0
\,,
\label{covariant conservation}
\end{equation}
it immediately follows that, in an epoch of constant $w$,
the energy density and pressure of a perfect fluid have
the following dependence on the scale factor,
\begin{equation}
\rho = \frac{p}{w} = \rho_0 \left(\frac{a_0}{a_{}} \right)^{2\epsilon} \ ,
\end{equation}
where $\rho(\eta_0)=\rho_0$, $a(\eta_0)=a_0$, and
\begin{equation}
\epsilon = \frac{3}{2}(1 + w) 
\,
\end{equation} 
is the so-called slow-roll parameter. A universe whose entire matter content
is dominated by one ideal fluid behaves in such a way that its slow-roll
parameter is constant,
\begin{equation}
\epsilon \equiv q+1 = 1 -\frac{\mathcal{H}'}{\mathcal{H}^2} = \mathrm{const} 
\,,
\end{equation}
where $q(t)=- a \ddot a/\dot a^2$ denotes the deceleration parameter.
The name of the $\epsilon$ parameter is inherited from the context of
inflationary models, but here it does not have to be small.
Some important values of this parameter are $\epsilon=2$ for
radiation-dominated Universe, $\epsilon=3/2$ for non-relativistic 
matter-dominated, and $\epsilon=0$ for an inflationary de Sitter 
universe. For FLRW spacetimes with constant $\epsilon$ parameter 
the Hubble rate and the scale factor are related as
\begin{equation}
\mathcal{H} = \mathcal{H}_0 \left( \frac{a_0}{a_{}} \right)^{\epsilon-1} \ ,
\end{equation}
and the explicit form of the scale factor is
\begin{equation}
a(\eta) = a_0 \Big[ 1 + (\epsilon-1)\mathcal{H}_0(\eta-\eta_0)
              \Big]^{\frac{1}{\epsilon-1}}
\, ,
\end{equation}
where $a(\eta_0)=a_0$, and $\mathcal{H}(\eta_0)=\mathcal{H}_0$.
In figure~\ref{figure 1:epsilon}
we show two examples of possible evolution of $\epsilon(t)$ throughout 
the history of the Universe: a period of inflation is followed by 
a period of radiation and matter era. On the left panel 
the transitions are quite sharp, {\it i.e.} the rate of transition
between different epochs is much larger than the expansion of the Universe,
while on the right panel the transition rate is quite mild. 
It is very hard to treat either of the two scenarios analytically,
so one has to resort to approximations. In this work we shall 
assume that transitions are sharp, and discuss in some detail 
the physical effects generated by a sharp transition.

\section{Scalar field on FLRW}
\label{Scalar field on FLRW}

The action of a massless, minimally coupled real scalar field living 
on a curved background is given by
\begin{equation}
S_\phi =  
- \frac{1}{2} \int d^Dx \sqrt{-g} g^{\mu\nu} \partial_{\mu}\phi\partial_\nu\phi
\ .
\label{scalar action} 
\end{equation}
Canonical quantization of this field on a FLRW background proceeds 
by first defining a canonical momentum,
\begin{equation}
\pi(x) = \frac{\delta S_{\phi}}{\delta \phi'}  = a^{D-2} \phi'(x) \ ,
\end{equation}
and then promoting the fields into operators, 
$\phi\rightarrow \hat \phi$, $\pi\rightarrow \hat \pi$, 
such that they satisfy the following canonical commutation relations,
\begin{align}
& [\hat{\phi}(\eta,\boldsymbol{x}),\hat{\pi}(\eta,\boldsymbol{x}')] = [\hat{\phi}(\eta,\boldsymbol{x}),a^{D-2} \hat{\phi}'(\eta,\boldsymbol{x}')] = i \delta^{D-1}(\boldsymbol{x}-\boldsymbol{x}') \ , \\
& [\hat{\phi}(\eta,\boldsymbol{x}),\hat{\phi}(\eta,\boldsymbol{x}')] = 0 = [\hat{\phi}'(\eta,\boldsymbol{x}),\hat{\phi}'(\eta,\boldsymbol{x}')]  
\, .
\end{align}
When these commutation relations are used in the Heisenberg equation, 
one obtains the equation of motion for the field operator
\begin{equation}
\hat{\phi}^{\prime\prime} + (D-2)\mathcal{H}\hat{\phi}^\prime 
 - \boldsymbol{\nabla}^2\hat{\phi} = 0 
\, .
\label{field EOM}
\end{equation}
Assuming a spatially homogeneous state $|\Omega\rangle$, 
the following decomposition of the field operator
(in Heisenberg picture) in Fourier modes is convenient, 
\begin{equation}
\hat{\phi}(\eta,\boldsymbol{x}) 
= a^{\frac{2-D}{2}} \int \frac{d^{D-1}k}{(2\pi)^{D-1}} 
\Big( {\rm e}^{i\boldsymbol{k}\cdot\boldsymbol{x}} u(k,\eta)
          \hat{a}(\boldsymbol{k})
 + {\rm e}^{-i\boldsymbol{k}\cdot\boldsymbol{x}} u^*(k,\eta) 
            \hat{a}^{\dag}(\boldsymbol{k}) 
 \Big)
\, ,
\end{equation}
where the assumed spatial homogeneity of the state dictates that 
the mode functions $u(k,\eta)$ and $u^*(k,\eta)$ depend on
the modulus $k=\|\boldsymbol{k}\,\|$ only. 
The annihilation operator, $\hat{a}$, and the creation operator, 
$\hat{a}^{\dag}$, satisfy 
\begin{equation}
[\hat{a}(\boldsymbol{k}),\hat{a}^{\dag}(\boldsymbol{k}')] 
  = (2\pi)^{D-1}\delta^{D-1}(\boldsymbol{k}-\boldsymbol{k}')
\, , 
\ \ \  [\hat{a}(\boldsymbol{k}),\hat{a}(\boldsymbol{k}')] 
       = [\hat{a}^\dag(\boldsymbol{k}),\hat{a}^{\dag}(\boldsymbol{k}')] = 0 
\, ,
\end{equation}
which fixes the Wronskian normalization of the mode functions,
\begin{equation} \label{Wronskian}
W[u(k,\eta),u^*(k,\eta)] = u(k,\eta) \, u'^*(k,\eta) - u^*(k,\eta) \, u'(k,\eta)  = i \ .
\end{equation}
The equation of motion for these mode functions is
\begin{equation}
u''(k,\eta) + \left[k^2 + f(\eta) \right] u(k,\eta) = 0 
\ ,
\label{mode function EOM}
\end{equation}
where $f$ is a time dependent `mass term',
\begin{equation}
f(\eta) = - \frac{D-2}{4}\Big{(}2\mathcal{H}'+(D-2)\mathcal{H}^2\Big{)} 
\ .
\label{f}
\end{equation}
The name of the creation and annihilation operator is justified by requiring that they construct the Fock space of states. In particular, 
the vacuum state $|\Omega\rangle$ is defined by
\begin{equation}
\hat{a}(\boldsymbol{k})|\Omega\rangle=0\,, 
\ \ \ \ (\forall \boldsymbol{k}) 
\, ,
\end{equation}
and all other states are built by acting on the vacuum state with 
creation operators. Notice that here we work in Heisenberg picture, in which  
time evolution is fully contained in the field operator, 
see Eqs.~(\ref{field EOM}) and~(\ref{mode function EOM}).

The physical choice of the mode function (the choice of the vacuum) is not as
straightforward on curved  backgrounds, as it is in Minkowski space,
where Poincar\'e symmetry can be used to uniquely fix the vacuum state. 
Here we concern ourselves only with this choice on constant
$\epsilon$ FLRW backgrounds. 
The usual choice on accelerating expanding backgrounds ($\epsilon<1$) 
is the Bunch-Davies vacuum. It is defined by requiring that
in the far past ($\eta\rightarrow-\infty$) the modes
behave as the positive-frequency conformal Minkowski space mode functions,
\begin{equation}\label{Conformal vacuum}
u(k,\eta) \;\;\stackrel{\eta\rightarrow-\infty}{\sim}\;\; 
    \frac{1}{\sqrt{2k}} \, {\rm e}^{-ik\eta}
\ ,
\end{equation}
Physically, this can be viewed 
as the requirement that the UV modes behave as those in conformally flat space,
which exhibits the Poincar\'e symmetry of Minkowski space,
since all the modes are more and more UV as one goes further to the past.
Moreover, one can show that in this adiabatic vacuum the energy density
functional, $E[\phi]/V=\int [d^3k/(2\pi)^3]\varepsilon(k,\eta)$ is minimized. 
To show that, let us for the moment assume that 
there is a mode mixing in the asymptotically distant past, 
such that instead of (\ref{Conformal vacuum}) we have, 
\begin{equation}\label{Conformal vacuum:2}
u(k,\eta) \;\;\stackrel{\eta\rightarrow-\infty}{\sim}\;\; 
    \frac{\alpha(k)}{\sqrt{2k}} \, {\rm e}^{-ik\eta}
    + \frac{\beta(k)}{\sqrt{2k}} \, {\rm e}^{ik\eta}
\,,\qquad  (|\alpha|^2-|\beta|^2=1)
\,. 
\end{equation}
It is then easily shown that the corresponding phase-space energy density
\begin{equation}
 \varepsilon(k,\eta)\simeq \frac{k}{2} \left( |\alpha|^2+|\beta|^2 \right)
\label{rho:modes}
\,,
\end{equation}
which is clearly minimized for $\beta=0$, giving a very strong argument 
in favor of the Bunch-Davies vacuum in the UV. The situation can, however, be 
very different for infrared modes in curved space-times such 
as expanding FLRW backgrounds, where adiabaticity assumed in 
deriving~(\ref{rho:modes}) is generally broken.

 The above definition of the usual Bunch-Davies state relies on the existence 
of an asymptotic past, which is the case in accelerating universes 
with $\epsilon<1$. Indeed, in this case asymptotic past means
more and more UV when compared to the curvature invariants. 
In decelerating universes (where $\epsilon>1$) however, 
asymptotic past does not exist since 
a decelerating universe starts at a finite time from a curvature singularity
(big bang). A simple fix of that problem is to consider 
the Bunch-Davies vacuum to be the UV state 
for which the mode functions are~(\ref{Conformal vacuum}) when 
$k/a\gg H$, {\it i.e.} physical momenta are large when compared to 
the space-time curvature, and this is what we shall assume here.
One more problem remains, and that is that the above procedure 
does not say anything about how to pick the state in the IR. 

 To address that question,
in this paper we extend the definition 
of the Bunch-Davies vacuum, to a 
{\it global Bunch-Davies vacuum} which we define as the analytic extension
of the state~(\ref{Conformal vacuum}).
Namely, while in the UV the distinction between positive and negative 
frequency modes is straightforward, it is not so in the IR. 
This analytic extension can be obtained either 
by analytically extending the asymptotic expansion for the mode functions 
that satisfy~(\ref{mode function EOM}) (which can be achieved for example 
by a Borel resummation), or by simply solving the equation of 
motion~(\ref{mode function EOM}) and choosing the solution such that 
it reduces to~(\ref{Conformal vacuum}) in the UV. Both methods will give 
the unique answer for the global Bunch-Davies vacuum state. 
When calculating the expectation values of our operators
below we shall always perform it with respect to
the global Bunch-Davies (BD) vacuum state.

\section{Energy-momentum tensor}
\label{Energy-momentum tensor}

The energy-momentum tensor operator of a massless, minimally coupled scalar field on a curved background is given by
\begin{equation}
\hat{T}_{\mu\nu}(x) 
= \left.\frac{-2}{\sqrt{-g}}\frac{\delta S_{\phi}}{\delta g^{\mu\nu}(x)}
  \right|_{\phi=\hat{\phi}} 
= \partial_\mu \hat{\phi} \, \partial_\nu\hat{\phi} 
 - \frac{1}{2}g_{\mu\nu} g^{\alpha\beta}\partial_\alpha\hat{\phi} \,
             \partial_\beta\hat{\phi} 
\ .
\end{equation}
Its expectation value on FLRW background, with respect to the global 
BD vacuum state $|\Omega\rangle$, which is diagonal, 
can be written in terms of integrals over the mode functions,
\begin{align}
 \langle\Omega |\hat{T}_{00}|\Omega\rangle
& = \frac{ a^{2-D}}{(4\pi)^{\frac{D-1}{2}}\Gamma\left( \frac{D-1}{2} \right)} 
       \int_0^\infty dk \, k^{D-2} \Bigg{\{} |u'|^2 + \left[k^2 
              + \frac{(D-2)^2}{4}\mathcal{H}^2\right]|u|^2 
                - \frac{D-2}{2}\frac{\partial}{\partial\eta}|u|^2 \Bigg{\}}
\ , 
\label{intT00} 
   \\
%
 \langle\Omega |\hat{T}_{ij}|\Omega\rangle 
&=\frac{\delta_{ij}a^{2-D}}{(4\pi)^{\frac{D-1}{2}}
         \Gamma\left( \frac{D-1}{2} \right)}
\label{intTij} \\
&\times \int_0^{\infty} dk\, k^{D-2} 
 \Bigg{\{} |u'|^2 + \left[ - \frac{D-3}{D-1}k^2
  + \frac{(D-2)^2}{4}\mathcal{H}^2 \right] |u|^2 
         - \frac{D-2}{2}\mathcal{H} \frac{\partial}{\partial\eta}|u|^2 
   \Bigg{\}} 
\ . \nonumber
\end{align}
The two integrals above, by themselves, generally exhibit quartic, quadratic, 
and logarithmic UV divergencies in $D=4$. 
Therefore, these integrals first need to be regularized in the UV, 
and the method we choose is dimensional regularization, which automatically 
subtracts power-law divergences. The remaining logarithmic divergence needs
to be absorbed into the higher derivative countertems which entails 
renormalization. The whole procedure of regularization and renormalization of 
the energy-momentum tensor on FLRW background is presented in Appendix A. 
In the end one associates finite values to the integrals above, $T_{00}^q$ 
and $T_{ij}^q$, which represent the physical expectation values of the energy
momentum tensor of the scalar field at one-loop.

 It is convenient to express the results in terms of energy density, $\rho_q$, 
and pressure, $p_q$,
\begin{equation}
\rho_q  = \frac{1}{a^2}T^q_{00} \ , \ \ \  \delta_{ij}p_q 
   = \frac{1}{a^2} T_{ij}^q  \ ,
\end{equation}
which can readily be compared to the corresponding values of the classical
fluids dictating the dynamics of FLRW background.

\section{Energy-momentum tensor on FLRW with constant $\epsilon$}
\label{Energy-momentum tensor on FLRW with constant epsilon}

In this section we examine the energy-momentum tensor of the scalar field
in a Bunch-Davies vacuum on three FLRW backgrounds of constant $\epsilon$ 
parameter, inflationary one with $\epsilon_I=-(D-4)/2$, 
radiation-dominated one with $\epsilon_R={D}/{2}$, and matter-dominated
one with $\epsilon_M=({D+2})/{4}$. Note that $D$-dependent $\epsilon$ 
parameters are just our prescription of how to extend their values in $D$=4 
to general $D$-dimensional FLRW backgrounds
(which are needed in order to employ dimensional regularization). 
This prescription of extension to $D$ dimensions is essentially arbitrary, 
and does not influence physical results in $D=4$. 
The motivation for this prescription is the that it keeps the mode function 
of the scalar field as simple in $D$ dimensions, as it was in $D$=4. 
One can see that from the equation of motion 
\eqref{mode function EOM} 
on constant $\epsilon$ background, written in terms of
the variable $y={k}/[{(1-\epsilon)\mathcal{H}}]$,
\begin{equation}
\left[ \frac{d^2}{d y^2} + 1 + \frac{\frac{1}{4}-\nu^2}{y^2} \right] u = 0 \ ,
\end{equation}
where
\begin{equation}
\nu^2 = \frac{1}{4} + \frac{(D\!-\!2)(D\!-\!2\epsilon)}{4(1\!-\!\epsilon)^2} 
       = \left(\frac{D\!-\!1\!-\!\epsilon}{2(1\!-\!\epsilon)}\right)^2 
\ .
\end{equation}
The properly normalized positive frequency solution of this equation is
\begin{equation}
u(k,\eta) = \sqrt{\frac{\pi}{4(1-\epsilon)\mathcal{H}}} \, 
  H_{\nu}^{(1)} \left( \frac{k}{(1-\epsilon)\mathcal{H}} \right) 
\ ,
\end{equation}
where $H_\nu^{(1)}$ is the Hankel function of the first kind. 
By requiring \eqref{epsD} we make the order of the Hankel function 
the same in arbitrary $D$ dimensions as in $D$=4. 
This is a matter of convenience, proving to be most practical 
in cases when $\nu=(2n+1)/2$, $n\in \mathbb{N}$ in $D=4$, 
in which case the Hankel function is just a finite sum. 
This is the case with all the mode functions we are working with in this paper.
This important simplification of mode functions is achieved by promoting 
$\epsilon$ to a $D$-dependent function $\epsilon_D$
according to the prescription
\begin{equation} \label{epsD}
\frac{(D-2)(D-2\epsilon_D)}{(1-\epsilon_D)^2} 
= \frac{4(2-\epsilon_4)}{(1-\epsilon_4)^2} 
\, .
\end{equation}

A brief comment regarding dimensional regularization is now warranted, 
when working with a $D$-dependent $\epsilon$ parameter. Although this
prescriptioni will change the finite terms in the UV one obtains from taking
the expectation value of the energy-momentum tensor operator, 
it will also change the finite
terms contributed from the counterterm which is to absorb the logarithmic 
divergence of the energy-momentum tensor. This will happen in such a way that 
these extra terms exactly cancel when the two are added together. 
A more explicit elaboration of this statement can be found at the end of
Appendix~A. As stated above, making $\epsilon$ $D$-dependent does not change
the physical result.

\subsection{Inflation}

On an inflationary background with
\begin{equation}
\epsilon_I = -\frac{D-4}{2} \xrightarrow{D\rightarrow 4} 0 \ ,
\end{equation}
the global Bunch-Davies mode functions are
\begin{equation}\label{infBD}
u_I(k,\eta) =\frac{1}{\sqrt{2k}} 
\left[ 1 + \frac{i(1\!-\!\epsilon_I)\mathcal{H}}{k} \right] 
\exp\left[\frac{ik}{(1\!-\!\epsilon_I)\mathcal{H}} \right] 
= \frac{1}{\sqrt{2k}} \left[ 1 + \frac{i(D\!-\!2)\mathcal{H}}{2k} \right] 
 \exp\left[ \frac{2ik}{(D\!-\!2)\mathcal{H}} \right] \ .
\end{equation}
Even though strictly speaking this mode function defines an IR singular
state on de Sitter, for the purposes of this paper this is irrelevant 
since the one-loop energy-momentum tensor we discuss here is
insensitive to the IR singular nature of the state as it is 
completely regular in the IR.  
Calculating the one-loop energy-momentum tensor amounts to going through
the procedure outlined in Appendix A. The final result is
\begin{align} \label{rhoINFL}
& \rho_q = - \frac{119\mathcal{H}^4}{960\pi^2a^4} 
\ , \ \ \ 
p_q =  \frac{119\mathcal{H}^4}{960\pi^2a^4} \ ,
\end{align}
from which we can infer the equation of state parameter,
\begin{equation}
w_q = -1 = w \ .
\end{equation}
From these results it follows that the scalar one-loop fluctuations
on de Sitter give a tiny contribution to the cosmological constant,
\begin{equation}
 \frac{\delta \Lambda}{\Lambda} = -\frac{119}{360\pi}
                     \bigg(\frac{\hbar H}{m_{\rm P}c^2}\bigg)^2
\label{delta Lambda: dS}
\end{equation}
where in the last step we explicitly introduced the dependences on 
$c$ and $\hbar$, and we defined the usual Planck mass, 
$m_{\rm P} = \sqrt{\hbar c/G_N}$.  
Although one should add this contribution to the cosmological term, 
on an exactly de Sitter background it is impossible to disentangle
the quantum contribution $\delta \Lambda$
from the classical contribution $\Lambda$. 

\subsection{Radiation}
\label{Radiation}

On a radiation dominated FLRW background with
\begin{equation}\label{rhoRAD}
\epsilon_R = \frac{D}{2} \xrightarrow{D\rightarrow 4} 2 \ ,
\end{equation}
the global Bunch-Davies mode function is
\begin{equation}\label{radBD}
u_R(k,\eta) = \frac{1}{\sqrt{2k}} 
\exp\left[ \frac{ik}{(1-\epsilon_R)\mathcal{H}} \right] 
= \frac{1}{\sqrt{2k}} \exp\left[ -\frac{2ik}{(D-2)\mathcal{H}} \right] \ .
\end{equation}
The energy density and pressure associated with this state are
\begin{align} \label{rhoRAD}
& \rho_q =  \frac{\mathcal{H}^4}{960\pi^2a^4} 
\ , \ \ \ 
p_q = \frac{\mathcal{H}^4}{576\pi^2a^4} \ ,
\end{align}
and the equation of state parameter is constant,
\begin{equation}
w_q = \frac{5}{3} > w = \frac{1}{3} 
\ ,
\end{equation}
which implies the scaling, $\rho_q\propto p_q\propto 1/a^8$.
We see that the contribution from the scalar fluctuations decays much faster
with the scale factor than the classical fluid driving the expansion, making 
them negligible at late times.

\subsection{Matter}

On a non-relativistic matter dominated FLRW background with
\begin{equation}
\epsilon_M = \frac{D+2}{4} \xrightarrow{D\rightarrow 4} \frac{3}{2} \ ,
\end{equation}
the global Bunch-Davies mode function is
\begin{equation} \label{matBD}
u_M(k,\eta) = \frac{1}{\sqrt{2k}} 
 \left[1+\frac{i(1\!-\!\epsilon_M)\mathcal{H}}{k} \right] 
 \exp \left[\frac{ik}{(1\!-\!\epsilon_M)\mathcal{H}} \right] 
= \frac{1}{\sqrt{2k}} \left[1\!-\!\frac{i(D\!-\!2)\mathcal{H}}{4k} \right] 
         \exp\left[ \frac{-4ik}{(D\!-\!2)\mathcal{H}} \right] \ .
\end{equation}
It turns out that on this background the global Bunch-Davies mode functions
are not a good physical choice of vacuum in the IR, since it gives rise to an
IR divergent one-loop energy-momentum tensor. 
One has to regularize this IR divergence in some way. Arguably the simplest way
is to introduce an IR cut-off $k_0$ in the integral defining 
the energy-momentum tensor, effectively regularizing the IR modes. 
This regularization can be achieved by placing  
the Universe in a box of a comoving size $\sim 2\pi/k_0$;
the discrete mode sums obtained in this way can be replaced 
by integrals plus corrections suppressed as powers of the IR cut-off scale 
$k_0$. The resulting one-loop energy density and pressure are then
\begin{align}
& \rho_q = \frac{9\mathcal{H}^4}{128\pi^2a^4} 
 \left[ \ln\left( \frac{\mu_f}{k_0}a \right) + \frac{61}{270} \right] 
\ , \\
& p_q =\frac{9\mathcal{H}^4}{128\pi^2a^4}
  \left[ \ln\left( \frac{\mu_f}{k_0}a \right) - \frac{29}{270} \right] 
\ ,  
\end{align}
where $\mu_f$ is a free constant that can in principle be determined by
measurements. The equation of state parameter is not constant,
\begin{equation}
w_q = \frac{\ln\left( \frac{\mu_f}{k_0}a \right) 
   - \frac{29}{270}}{\ln\left( \frac{\mu_f}{k_0}a \right) + \frac{61}{270}} 
\ .
\end{equation}
If a proper care is taken to define the initial condition, 
then $w_q>0=w$, meaning that that quantum contribution is negligible.
The drawback of this regularization is that it is not valid for all times, 
since the Hubble rate is a time-evolving scale. At the point when 
$k_0 =  \mathcal{H}(\eta)$ the regularization breaks down because 
the cut-off becomes UV instead of IR. Nevertheless, the above expressions 
for the energy density and pressure are valid for a limited time period. 
A more proper way to define the vacuum on this background would be to temper
the IR behavior of the mode functions by introducing 
appropriate Bogolyubov coefficients, see 
{\it e.g.}~\cite{Janssen:2009nz,Janssen:2008dw}.

\section{Energy-momentum on FLRW with evolving $\epsilon$}
\label{Energy-momentum on FLRW with evolving epsilon}

 From the three examples of the preceding section we see that,
since $w_q \ge w$, the global Bunch-Davies state on a period of constant
$\epsilon$ does not lead to any significant backreaction. 
But we know that the time evolution of $\epsilon$ parameter such as 
illustrated on figure~\ref{figure 1:epsilon} leads to 
the quantum state evolving away from the Bunch-Davies vacuum state
of the particular era, which is often characterized as particle production. 
That also leads to the different contributions to the energy density
and pressure. We study these contributions in this section.

The transitions between different periods of the expansion of the Universe
proceed in a smooth manner. These transitions can be described by 
a time-evolving $\epsilon(\eta)$, which can in principle be determined from
the known classical matter content (in the form of perfect fluids) driving
the expansion and their interactions. 
The practical problem is that there are very few
$\epsilon(\eta)$ for which  the equation of motion for the mode
function \eqref{mode function EOM} admits analytically tractable solutions.
For an example where such an interpolating solution can be constructed see
Ref.~\cite{Koivisto:2010pj}.
Even if one can find a solution for the mode function, there is
still the task of performing all the integrations 
in~(\ref{intT00}--\ref{intTij}) to get the energy-momentum tensor. 
Unless one wants to study the contribution
of the quantum fluctuations numerically, approximating the full behavior 
of $\epsilon(\eta)$ is a necessity. 

It is well known that in the case of constant $\epsilon$ one can perform all
calculations in the closed form. For this reason we model our FLRW background
to have constant $\epsilon$ parameters during some time intervals, 
and the transitions between different eras are assumed to be instantaneous.
By assuming this for the background, we limit the regions where physical
predictions for energy density and pressure of the quantum fluctuations
can be made. Namely, we do not expect to describe well enough the behavior
close to the point of transition, but might be able to determine 
correctly the late time behavior, if the details of the transition
get lost with time. 
The main purpose of this section is to discuss in some detail
how one can make accurate estimates of the one-loop energy momentum
tensor at times sufficiently long after the transition.

\subsection{One sudden transition}
\label{One sudden transition}

First, we will address the issues arising from the scale factor of 
a FLRW background having a discontinuity in its $n$th derivative, 
$n\ge2$, at some $\eta_0$ (sudden transitions in $\epsilon$ parameter described
above correspond to $a^{\prime\prime}(\eta)$ having a discontinuity at $\eta_0$). 
The discontinuity in the background spoils adiabaticity of 
the mode function in the UV (for momenta $k>\mu\gg \mathcal{H}$), 
on which we relied in Appendix A to determine the UV divergent structure
of $\langle T_{\mu\nu} \rangle$. This discontinuity will induce 
a mixing of positive and negative frequency modes in the UV, 
which will not be adiabatically suppressed. Here we derive what this mixing is,
how it depends on the order of continuity (or smoothness) of the background, 
and its influence on the UV part of the energy-momentum tensor.

For $\eta<\eta_0$ we assume that the full mode function, $U(k,\eta)$, 
in the UV is the positive-frequency global Bunch-Davies one, 
$u(k,\eta)$, given by \eqref{symb asympt} in Appendix A. A discontinuity
in the background at $\eta_0$ will spoil adiabaticity of the mode function.
The full mode function will still evolve adiabatically after $\eta_0$, 
but it will no longer be a pure positive-frequency function, 
as the discontinuity in the background will induce some mode mixing. 
The order of this mixing will depend on the order of discontinuity. 
After $\eta_0$, the full mode function will be a linear superposition of
the positive and negative frequency modes,
\begin{equation}
U(k,\eta) = \alpha(k)u(k,\eta) + \beta(k)u^*(k,\eta) \ ,
\end{equation}
where $u(k,\eta)$ is the UV positive frequency function \eqref{symb asympt} 
on a smooth FLRW background for $\eta>\eta_0$, and $\alpha(k)$ and 
$\beta(k)$ are the Bogolyubov coefficients representing mode mixing, 
which have to satisfy
\begin{equation}\label{BogUnit}
|\alpha(k)|^2 - |\beta(k)|^2 = 1 
\ ,
\end{equation}
because of the Wronskian normalization \eqref{Wronskian} being dictated 
by canonical quantization.
These coefficients are the solution of the continuity conditions 
on the mode function
\begin{equation} \label{match}
U(k,\eta_0^-) = U(k,\eta_0^+) \ , \ \ \ U'(k,\eta_0^-) = U'(k,\eta_0^+) \ ,
\end{equation}
where the superscripts on the time arguments signify from which direction 
the limit to $\eta_0$ is taken, 
$\eta^\pm_0={\rm lim}_{\delta_\eta\rightarrow 0}[\eta_0\pm\delta_\eta]$.
From here, using \eqref{symb asympt} and \eqref{F1}-\eqref{F4}, 
it is straightforward to derive the expressions for $\beta(k)$ in the UV,
\begin{equation} \label{BetaUV}
\beta(k) = e^{-2ik\eta_0} \left[ \frac{B_2}{k^2} + \frac{iB_3}{k^3} + \frac{B_4}{k^4} + \frac{iB_5}{k^5} \right] + \mathcal{O}(k^{-6}) \ ,
\end{equation}
where
\begin{align}
B_2 ={}& \frac{1}{4} \Big{[} f(\eta_0^+) - f(\eta_0^-) \Big{]} \ , \label{B2} \\
B_3 ={}& - \frac{1}{8} \Big{[} f'(\eta_0^+) - f'(\eta_0^-) \Big{]} \ , \\
B_4 ={}& - \frac{1}{16} \Big{[} f''(\eta_0^+) - f''(\eta_0^-) + 2\Big{(}f^2(\eta_0^+)-f^2(\eta_0^-) \Big{)} \Big{]} \ , \\
B_5 ={}& \frac{1}{32} \Big{[} f'''(\eta_0^+) - f'''(\eta_0^-) + 5\Big{(} f'(\eta_0^+) f(\eta_0^+) - f'(\eta_0^-) f(\eta_0^-) \Big{)} + \nonumber \\
&+ f'(\eta_0^+) f(\eta_0^-) - f'(\eta_0^-) f(\eta_0^+) \Big{]} \label{B5}  \ ,
\end{align}
and $f(\eta)$ is defined in \eqref{f}.
Calculating $\alpha(k)$ is also straightforward, but for our purposes it suffices to notice that its leading order in UV is always 1, 
and that it can be written as
\begin{equation}
\alpha(k) = 1 + \frac{iA_1}{k} + \frac{A_2}{k^2} + \mathcal{O}(k^{-3}) \ ,
\label{alpha: adiabatic expansion}
\end{equation}
where $A_n$ are real coefficients whose exact value is not important
for the present discussion. From \eqref{BetaUV} and \eqref{B2}-\eqref{B5} 
it is clear how the order of smoothness of the sudden transition relates 
to the leading order of the Bogolyubov coefficient $\beta$ in the UV, 
since $f(\eta)$ contains the second time derivative of the scale factor. 
If the FLRW background has a discontinuity in the $n$th derivative 
of its scale factor, then the leading order in $\beta(k)$ in the UV 
will be ${\cal O}(k^{-n})$.

Noting that the squares of the full mode function to be integrated over
can be written as
\begin{align}
 |U|^2 &= |u|^2 + 2|\beta|^2|u|^2 + \alpha\beta^*u^2 + \alpha^*\beta u^{*2} 
\ , \\
 |U'|^2 &= |u'|^2 + 2|\beta|^2|u'|^2 + \alpha\beta^*u'^2 
   + \alpha^*\beta u'^{*2} \ , 
\\
 \frac{\partial}{\partial\eta}|U|^2
 &= \frac{\partial}{\partial\eta}|u|^2 
 + 2|\beta|^2\frac{\partial}{\partial\eta}|u|^2 
+ \alpha\beta^*\frac{\partial}{\partial\eta}u^2 
+ \alpha^*\beta \frac{\partial}{\partial\eta}u^{*2} 
\ ,
\end{align}
where we used \eqref{BogUnit}, we can split the UV part of the energy-momentum
tensor into three parts,
\begin{equation} \label{sep}
\langle\Omega | \hat{T}_{\mu\nu} |\Omega\rangle_{UV} 
= \langle T_{\mu\nu} \rangle_{UV}^{BD} 
+ \langle T_{\mu\nu}\rangle_{UV}^{\beta} 
+ \langle T_{\mu\nu}\rangle_{UV}^{\alpha\beta} 
\ ,
\end{equation}
to be defined below. The first part, $\langle T_{\mu\nu}\rangle_{UV}^{BD}$,
is the part that one also has for a pure positive-frequency Bunch-Davies
mode function, as described in Appendix~A. 
We do not need to calculate this contribution again, 
the results carry over completely. The remaining two terms are a consequence
of the mode mixing induced by the background discontinuity.
These terms would be zero in the UV if the background was completely smooth. 
Therefore, we have to examine them more closely.

The middle term in \eqref{sep} is
\begin{align}
\langle T_{00}\rangle_{UV}^{\beta} 
={}& \frac{2a^{2-D}}{(4\pi)^{\frac{D-1}{2}}\Gamma\left( \frac{D-1}{2} \right)}
\int_{\mu}^{\infty} dk\, k^{D-2} |\beta(k)|^2 \left\{ |u|^2 
  + \left[ k^2 + \frac{(D-2)^2}{4}\mathcal{H}^2 \right] |u|^2 
  - \mathcal{H} \frac{\partial}{\partial\eta} \right\} 
\nonumber \\
={}& \frac{2a^{2-D}B_2^2}{(4\pi)^{\frac{D-1}{2}}
\Gamma\left( \frac{D-1}{2} \right)} \int_{\mu}^{\infty} dk\, k^{D-5}  
    + \mathcal{O}(\mu^{-2}) 
\ , 
\label{Tbeta00} 
\\
\langle T_{ij}\rangle_{UV}^{\beta} 
={}& \frac{2\delta_{ij}a^{2-D}}{(4\pi)^{\frac{D-1}{2}}
      \Gamma\left( \frac{D-1}{2} \right)} 
           \int_{\mu}^{\infty} dk\, k^{D-2} |\beta(k)|^2 
      \left\{ |u|^2 + \left[ -\frac{D-3}{D-1}k^2 
       + \frac{(D-2)^2}{4}\mathcal{H}^2 \right] |u|^2 
        - \mathcal{H} \frac{\partial}{\partial\eta} \right\} 
\nonumber \\
={}& \frac{2a^{2-D}B_2^2}{(D-1)(4\pi)^{\frac{D-1}{2}}
       \Gamma\left( \frac{D-1}{2} \right)} 
 \int_{\mu}^{\infty} dk\, k^{D-5} + \mathcal{O}(\mu^{-2}) 
\label{Tbetaij} 
\ ,
\end{align}
where we dropped the $F_i$ contributions defined in~(\ref{symb asympt}),
which is allowed in the UV because $|\beta|^2\propto k^{-4}$.
The integrals~(\ref{Tbeta00}--\ref{Tbetaij})
are obviously divergent for all times after the matching, 
provided $B_2\ne 0$. The last term in \eqref{sep} can be immediately evaluated
in $D=4$,
\begin{align}
\langle T_{00} \rangle_{UV}^{\alpha\beta}
 ={}& \frac{1}{4\pi^2a^2} \int_{\mu}^{\infty} dk \, k^2 \alpha(k)\beta^*(k)
        \left[ u'^2 + \left( k^2 + \mathcal{H}^2 \right) u^2 
          - \mathcal{H} \frac{\partial}{\partial\eta} u^2 \right] + {\rm c.c.}
\nonumber \\
={}& \frac{1}{4\pi^2a^2} \Big{\{} 2B_2 \mathcal{I}_0^{\mu} 
+ \Big{[} B_2(\mathcal{H}^2 - 4F_1\mathcal{H} + 2F_1' 
- 2A_1\mathcal{H}) -2 B_3\mathcal{H} \Big{]} \mathcal{I}_1^\mu \Big{\}} 
+ \mathcal{O}(\mu^{-1}) \ , \\
\langle T_{ij} \rangle_{UV}^{\alpha\beta} ={}& \frac{\delta_{ij}}{4\pi^2a^2} 
\int_\mu^{\infty} dk\, k^2 \alpha(k) \beta^*(k) \left[ u'^2 
+ \left( -\frac{2k^2}{3} + \mathcal{H}^2 \right) u^2 
- \mathcal{H} \frac{\partial}{\partial\eta}u^2 \right] + {\rm c.c.}
\nonumber \\
={}& \frac{\delta_{ij}}{4\pi^2a^2} 
 \bigg\{  
     -\frac53 B_2 {\cal I}_{-1}^\mu
     +\bigg[
         B_2\Big(2{\cal H}-\frac{10}{3}F_1-\frac53A_1\Big)
              -\frac53B_3
      \bigg]{\cal I}_0^\mu
\nonumber\\
    & +\bigg[
         B_2\Big(-\frac{10}{3}F_2+2F_1^\prime+\frac53F_1^2
           -4{\cal H}F_1+{\cal H}^2-\frac{5}{3}A_2+\frac{10}3A_1F_1
                -2{\cal H}A_1\Big)
\nonumber\\
    & + B_3\Big(-2{\cal H}+\frac{10}{3}F_1+\frac53A_1
           \Big)
           -\frac{5}{3}B_4
      \bigg]{\cal I}_1^\mu
 \bigg\} 
  + \mathcal{O}(\mu^{-1}) 
\ ,
\end{align}
where the integrals are
\begin{align}
\mathcal{I}_{-1}^{\mu} 
 ={}& \int_\mu^{\infty} dk\, k \cos(k\Delta) 
= - \frac{1}{\Delta^2} \Big{[} \cos(\mu\Delta) 
+ \mu\Delta\sin(\mu\Delta) \Big{]} 
\label{Imu-1} 
\ , \\
\mathcal{I}_0^\mu ={}& \int_\mu^{\infty} dk \sin(k\Delta) 
 = \frac{\cos(\mu\Delta)}{\Delta} 
\ , \label{Imu0} 
\\
\mathcal{I}_1^\mu ={}& \int_\mu^\infty \frac{dk}{k} \cos(k\Delta)
 = - \mathrm{Ci}(\mu\Delta)
 \label{Imu1} 
\ ,
\end{align}
where $\Delta=2(\eta-\eta_0)$, and $\mathrm{Ci}$ is the cosine integral
function. The contributions above are finite for all $\eta>\eta_0$, 
but in the limit $\eta\rightarrow\eta_0$ they diverge 
(hence we call them boundary-divergent terms). 
The two types of divergences encountered here need to be regulated somehow
in order to obtain a physical answer for the energy-momentum tensor.

Keeping in mind the more physical picture of smooth backgrounds, 
we would not expect the transition between the periods of 
two different constant $\epsilon$'s to introduce any new divergences. 
But, when $B_2,\ne 0$, as it is in this paper, 
this is exactly what happens, a new logarithmic divergence appears, 
together with boundary divergences, which are present for any 
$B_2, B_3, B_4\ne 0$. By making the scale factor more continuous 
($d^4a/d\eta^4$ continuous), all these divergences disappear. 
Led by the knowledge of how the UV modes behave on smooth backgrounds, 
where $\beta$ is zero to all adiabatic orders  \cite{Parker1969}, 
we propose a way to regulate these divergences. 
We append an exponential damping term for $\beta$,
mimicking this non-adiabatic suppression, 
$\beta\rightarrow\beta e^{-\tau k}$. The parameter $\tau$ introduced
is a toy model for some very small, but finite time scale of the transition, 
$\tau\ll \mathcal{H}_0^{-1}$~\footnote{A better way of regularizing 
the transition from say $\epsilon_1$ to $\epsilon_2$ is to change
$f(\eta)_{\rm sudden}= f_1(\eta)+[f_2(\eta)-f_1(\eta)]\Theta(\eta-\eta_0)$
defined in Eq.~(\ref{f}) 
to a smooth function, 
$f(\eta)_{\rm smoooth}= f_1(\eta)+[f_2(\eta)-f_1(\eta)]
       {\rm tanh}[(\eta-\eta_0)/\tau]$, where $f_1$ and $f_2$ 
correspond to space-time with $\epsilon_1={\rm constant}$ 
and $\epsilon_2={\rm constant}$, respectively.
The problem is that we do not know how to solve exactly the resulting mode
equation, and to push this procedure through
suitable approximations ought to be applied.
}. 
This clearly regulates all the integrals above. But, if we modify $\beta$, 
we should modify $\alpha$ accordingly, so as to satisfy relation
\eqref{BogUnit}, as required by canonical quantization. 
We use that very relation to define the modification for the modulus 
of $\alpha$,
\begin{equation}
|\alpha| = \sqrt{1+|\beta|^2} \rightarrow \sqrt{1+|\beta|^2e^{-2\tau k}} \ ,
\end{equation}
so that we modify the whole coefficient as
\begin{equation} \label{mod alpha}
\alpha \rightarrow \sqrt{\frac{1+|\beta|^2e^{-2\tau k}}{1+|\beta|^2}} \, \alpha \ .
\end{equation}

The regulating procedure for transitions defined here renders the divergent, 
and boundary-divergent integrals everywhere finite. 
The integral in \eqref{Tbeta00} and \eqref{Tbetaij} becomes 
(now evaluated in $D=4$ right away)
\begin{equation}
\int_\mu^{\infty} \frac{dk}{k} \rightarrow \int_\mu^{\infty} 
\frac{dk}{k} e^{-2\tau k} = \Gamma(0,2\mu\tau) \ , 
\end{equation}
where $\Gamma(s,x)$ is the upper incomplete gamma function. 
The integrals \eqref{Imu-1}-\eqref{Imu1} after regularization become
(where we can expand the modification of $\alpha$ in the UV),
\begin{align}
\mathcal{I}_{-1}^\mu \rightarrow{}& \int_\mu^\infty dk\,  
{\rm e}^{-\tau k} k \cos(k\Delta) + \mathcal{O}(e^{-\tau\mu} \mu^{-1})  
\nonumber \\
={}& \frac{{\rm e}^{-\tau\mu}}{(\Delta^2+\tau^2)^2} \Big{\{} \cos(\mu\Delta) 
\Big{[} -\Delta^2+\tau^2 + \mu\tau(\Delta^2+\tau^2) \Big{]} 
- \Delta \sin(\mu\Delta) \Big{[} 2\tau + \mu(\Delta^2+\tau^2) \Big]\Big\}
\ , \\
\mathcal{I}_0^{\mu} \rightarrow{}& \int_\mu^{\infty} dk\, 
{\rm e}^{-\tau k} \sin(k\Delta) + \mathcal{O}(e^{-\mu\tau}\mu^{-1}) 
= \frac{{\rm e}^{-\tau\mu}}{\Delta^2+\tau^2} 
\Big[\Delta \cos(\mu\Delta) + \tau \sin(\mu\Delta) \Big]
\ , \\
\mathcal{I}_1^{\mu} \rightarrow{}& \int_\mu^{\infty} dk 
\frac{{\rm e}^{-\tau k}}{k} \cos(k\Delta) + \mathcal{O}(e^{-\mu\tau}\mu^{-1}) 
= - \frac{1}{2} \Big{[} \mathrm{Ei}(-\mu(\tau-i\Delta)) 
 + \mathrm{Ei}(-\mu(\tau+i\Delta))  \Big{]} 
\ ,
\end{align}
where $\mathrm{Ei}$ is the exponential integral function.

As for the IR, this regulating procedure also alters the IR, 
but only slightly, since modifying factors for $\alpha$ and $\beta$ quickly
reduce to $1$ in the IR. 
Therefore, the property found in \cite{FP}, that an IR-finite state will not
develop any IR divergences in time still holds. On the practical side, 
we will not be able to calculate the IR contribution using the full suppression
coefficient for $\alpha$ \eqref{mod alpha}. What will suffice in calculating
the energy-momentum tensor integrals is just taking this coefficient
to be $1$. The reason for this is that in the end it can only produce terms
of order $\mathcal{\tau}$ which are regular everywhere, 
and these terms we will drop in the limit when $\tau\rightarrow 0$. 
As a consequence, we will obtain
an approximate answer for the energy density and pressure, which will
satisfy the conservation equation~(\ref{covariant conservation}) 
approximately, to order $\mathcal{O}(\tau)$ corrections.

To conclude this general treatment of discontinuities in the FLRW background, 
we note that if one wants to obtain the energy momentum tensor of
the  quantum scalar field living on such a background that is fully regular
everywhere, without the need for a posteriori regularization presented here, 
one needs to have scale factor which is at least of a smoothness
class $\mathcal{C}^4$.

\subsection{Matching inflation to radiation}
\label{Matching inflation to radiation}

In this section we demonstrate the procedure detailed in the previous section
on a specific example simple enough to keep everything transparent: a sudden 
transition from an inflationary period to a radiation era.

Let the FLRW background be specified by 
$\epsilon_I = -(D\!-\!4)/2$ for $\eta<\eta_0$, and by 
$\epsilon_R = {D}/{2}$, for $\eta>\eta_0$, where $\eta_0$ is some time
of matching, and $\mathcal{H}_0$ is the conformal Hubble rate at that time. 
For $\eta<\eta_0$ the full mode function of the scalar field is taken to be
a global Bunch-Davies one \eqref{infBD},
\begin{equation}
U(k,\eta<\eta_0) = u_I(k,\eta) 
= \frac{1}{\sqrt{2k}} \left[ 1 + \frac{i(D-2)\mathcal{H}}{2k} \right] 
\exp\left[ \frac{2ik}{(D-2)\mathcal{H}} \right] 
\ ,
\end{equation}
and the contribution to the energy-density and pressure is given 
by \eqref{rhoINFL}.
After the matching, the full mode function will no longer be
a positive frequency mode function \eqref{radBD}, but instead it will be
a linear combination,
\begin{equation}
U(k,\eta>\eta_0) = \alpha_R(k) u_R(k,\eta) + \beta_R(k) u_R^*(k,\eta) \ ,
\end{equation}
where $u_R(k,\eta)$ is a radiation period global Bunch-Davies mode function
\eqref{radBD}, and $\alpha_R(k)$, $\beta_R(k)$ are Bogolyubov coefficients,
determined from the matching condition \eqref{match},
\begin{align}
& \alpha_R(k) = \left[ 1 + \frac{i\mathcal{H}_0}{k} 
- \frac{\mathcal{H}_0^2}{2k^2} \right] e^{\frac{2ik}{\mathcal{H}_0}} 
\ , \ \ \ 
\beta_R(k) = \frac{\mathcal{H}_0^2}{2k^2} \ ,
\end{align}
where we have used the conclusion from the previous section that we can
calculate with Bogolyubov coefficients immediately in $D=4$. 
Next, we separate the contributions to the one-loop energy-momentum tensor
as in \eqref{sep}, the first contribution being just 
a Bunch-Davies one \eqref{rhoRAD}.

The second part we calculate immediately in $D=4$, with an exponential
suppression as described in the previous section, where we assume 
$\tau\ll \mathcal{H}_0^{-1}$,
\begin{align}
\langle T_{00}\rangle^{\beta}={}& \frac{1}{2\pi^2a^2} 
\int_{k_0}^{\infty} dk\, k^2 |\beta_R|^2 \left[ k
+ \frac{\mathcal{H}}{2k^2} \right] e^{-2\tau k} 
= \frac{\mathcal{H}_0^4}{8\pi^2a^2} \left[ -\gamma_E - \ln(2 k_0\tau) 
        + \frac{\mathcal{H}^2}{4 k_0^2} \right] +{\cal O}(k_0)
\ , \\
\langle T_{ij} \rangle^{\beta} ={}& \frac{\delta_{ij}}{2\pi^2a^2}
 \int_{k_0}^{\infty} dk\, k^2 |\beta_R|^2 \left[ \frac{k}{3} 
 + \frac{\mathcal{H}}{2k^2} \right] e^{-2\tau k} 
= \frac{\delta_{ij}\mathcal{H}_0^4}{8\pi^2a^2} 
\left[ - \frac{\gamma_E}{3} - \frac{1}{3}\ln(2 k_0\tau) 
  + \frac{\mathcal{H}^2}{4 k_0^2} \right] +{\cal O}(k_0)
\ ,
\end{align}
where we have used 
$\Gamma(0,k_0\tau) = -\gamma_E - \ln(k_0\tau) + \mathcal{O}(k_0)$, 
and $\gamma_E$ is the Euler-Mascheroni constant constant. 
Notice that we have introduced an IR cut-off here, since the integrals are
IR divergent. This does not signal an IR divergence of
the whole energy-momentum tensor; it is just a consequence of
our way of splitting the contributions. 
Individual integrals might be IR divergent,
but these divergences cancel in the full answer, 
when all the contributions are added together. In fact, since we have started
with an IR-finite state (in inflation), we are guaranteed that that
state will not develop any IR divergences \cite{FP} as it evolves.

The last remaining contribution to the energy-momentum tensor is (also calculated directly in $D=4$)
\begin{align}
\langle T_{00} \rangle^{\alpha\beta} ={}& \frac{1}{4\pi^2a^2} \int_{k_0}^{\infty} dk\, k^2 e^{-\tau k} \left[ \alpha_R \beta_R^* \left( i\mathcal{H} + \frac{\mathcal{H}^2}{2k} \right) e^{-\frac{2ik}{\mathcal{H}}} + \alpha_R^* \beta_R \left( -i\mathcal{H} + \frac{\mathcal{H}^2}{2k} \right) e^{\frac{2ik}{\mathcal{H}}} \right] = \nonumber \\
={}&\frac{\mathcal{H}_0^2}{8\pi^2a^2} \left[ -\frac{\mathcal{H}_0^2\mathcal{H}^2}{4k_0^2} + \frac{2\mathcal{H} \Delta}{\Delta^2+\tau^2} + \frac{\mathcal{H}_0^2}{2}\ln\big{[} k_0^2(\Delta^2+\tau^2) \big{]} + \gamma_E\mathcal{H}_0^2 - \frac{1}{2}(\mathcal{H}_0-\mathcal{H})^2 \right] \ , \\
\langle T_{ij} \rangle^{\alpha\beta} ={}& \frac{\delta_{ij}}{4\pi^2a^2} \int_{k_0}^{\mu} dk\, k^2 e^{-\tau k} \left[ \alpha_R\beta_R^* \left( -\frac{2k}{3} + i\mathcal{H} + \frac{\mathcal{H}^2}{2k} \right) e^{-\frac{2ik}{\mathcal{H}}} + \alpha_R^*\beta_R \left( -\frac{2k}{3} - i\mathcal{H} + \frac{\mathcal{H}^2}{2k} \right) e^{\frac{2ik}{\mathcal{H}}}  \right] = \nonumber \\
={}&  \frac{\delta_{ij}\mathcal{H}_0^2}{8\pi^2a^2} \Bigg{[} -\frac{\mathcal{H}_0^2\mathcal{H}^2}{4k_0^2} + \frac{4}{3} \frac{\Delta^2-\tau^2}{(\Delta^2+\tau^2)^2} + \frac{2\Delta}{\Delta^2+\tau^2}\left( \mathcal{H} - \frac{2\mathcal{H}_0}{3} \right) + \nonumber \\
&+ \frac{\mathcal{H}_0^2}{6} \ln\big{[} k_0^2 (\Delta^2+\tau^2) \big{]} + \frac{\gamma_E\mathcal{H}_0^2}{3} - \frac{1}{2}(\mathcal{H}_0-\mathcal{H})^2 \Bigg{]} \ ,
\end{align}
where $\Delta=\frac{2(\mathcal{H}_0-\mathcal{H})}{\mathcal{H}_0\mathcal{H}}$, and where we used the integrals from Appendix~B.

Adding up all three contributions yields the final answer for energy density and pressure,
\begin{align}
\rho_q ={}& \frac{1}{8\pi^2a^4} 
\left[\frac{2\mathcal{H}_0^2\mathcal{H}\Delta}{\Delta^2+\tau^2} 
 + \frac{\mathcal{H}_0^4}{2} \ln\left( \frac{\Delta^2+\tau^2}{4\tau^2} \right) 
- \frac{\mathcal{H}_0^2}{2}(\mathcal{H}_0-\mathcal{H})^2 
 + \frac{\mathcal{H}^4}{120} \right] 
\ , \\
p_q ={}& \frac{1}{8\pi^2a^4} \Bigg{[} \frac{4\mathcal{H}_0^2}{3} \frac{\Delta^2
-\tau^2}{(\Delta^2+\tau^2)^2} 
+ \frac{2\mathcal{H}_0^2\Delta}{\Delta^2+\tau^2}\left( \mathcal{H}
-\frac{2\mathcal{H}_0}{3} \right) 
+ \frac{\mathcal{H}_0^4}{6}\ln\left(\frac{\Delta^2+\tau^2}{4\tau^2} \right) 
\nonumber \\
& - \frac{\mathcal{H}_0^2}{2}(\mathcal{H}_0-\mathcal{H})^2 
+\frac{\mathcal{H}^4}{72} \Bigg{]} \ .
\end{align}
These expressions for the energy density and pressure are completely regular
after the matching, and they satisfy the conservation 
equation~(\ref{covariant conservation}) to order $\mathcal{O}(\tau^2)$.

We can see that at late times, when $\mathcal{H}\rightarrow 0$ and
$\Delta\rightarrow \infty$, the contributions containing the parameter $\tau$ 
become negligible (except in the denominator of the logarithm),
\begin{align}
\rho_q ={}& \frac{1}{8\pi^2a^4} 
\left[ \frac{\mathcal{H}_0^3\mathcal{H}^2}{\mathcal{H}_0-\mathcal{H}} 
+ \mathcal{H}_0^4 \ln\left( \frac{\mathcal{H}_0-\mathcal{H}}
{\tau\mathcal{H}_0\mathcal{H}} \right) 
- \frac{\mathcal{H}_0^2}{2}(\mathcal{H}_0-\mathcal{H})^2 
+ \frac{\mathcal{H}^4}{120} \right] 
\ , \\
p_q ={}& \frac{1}{8\pi^2a^4} \Bigg{[} 
\frac{\mathcal{H}_0^4\mathcal{H}^2}{3(\mathcal{H}_0-\mathcal{H})^2} 
+ \frac{\mathcal{H}_0^3\mathcal{H}}{\mathcal{H}_0-\mathcal{H}}
\left( \mathcal{H}-\frac{2\mathcal{H}_0}{3} \right) 
+ \frac{\mathcal{H}_0^4}{3} \ln\left( \frac{\mathcal{H}_0
-\mathcal{H}}{\tau \mathcal{H}_0\mathcal{H}} \right) 
\nonumber \\
& - \frac{\mathcal{H}_0^2}{2}(\mathcal{H}_0-\mathcal{H})^2 
+\frac{\mathcal{H}^4}{72} \Bigg{]} \ .
\end{align}
The late time behavior does not depend on regularization
of the boundary divergent terms in $\langle T_{\mu\nu} \rangle^{\alpha\beta}$. 
It only depends on regularization of the logarithmic divergence in
$\langle T_{\mu\nu} \rangle^{\beta}$, which is understandable, 
since this divergence persists for all times if left unregulated.

Finally, we can discuss the dominant behavior at late times
($\eta\gg \eta_0$). Since $\mathcal{H}\sim a^{-1}$, and $a \sim \eta$ for
radiation dominated period, the dominant contribution as 
$\eta\rightarrow\infty$ is
\begin{equation}
\rho_q = - \frac{\mathcal{H}_0^4}{8\pi^2a^4} 
   \left[\ln\left( \frac{\tau\mathcal{H}}{2} \right) +\frac12\right]
      +{\cal O}(a^{-5})
\,,\qquad
p_q = - \frac{\mathcal{H}_0^4}{24\pi^2a^4} 
          \left[\ln \left( \frac{\tau\mathcal{H}}{2} \right)+ \frac32\right]
      +{\cal O}(a^{-5})
\ .
\end{equation}
The corresponding equation of state parameter at late times is then
\begin{equation}
w_q \simeq \frac{1}{3} = w 
\ .
\end{equation}
Although the equation of state parameter is constant at late times, 
the quantum contribution does not behave like a classical radiation fluid, 
since, on top of the radiation-like scaling $a^{-4}$,
there is a logarithmic enhancement factor $\ln(a)$ 
in both the energy density and pressure. 
This means that the quantum contribution to the energy density and pressure
will eventually become comparable to the classical contribution, 
and will alter the expansion in a way we cannot predict without
self-consistently solving the backreaction problem.

 Nevertheless, if the Universe continues expanding as radiation dominated
we can roughly estimate when the quantum contribution would become comparable. 
To get that estimate let us reinsert units in the dominant late-time behavior
and rewrite it in terms of physical Hubble rate, $H=\mathcal{H}/a$,
\begin{eqnarray}
\rho_q &=& \frac{\hbar H_0^4}{8\pi^2c^3} \left( \frac{a_0}{a} \right)^4 
  \left\{\ln \left[\frac{2}{(a_0\tau) H_0} \left( \frac{a}{a_0} \right) \right]
   +\frac12\right\} +{\cal O}(a^{-5})
\label{rhoq and pq:radiation}
\\
 p_q &=& \frac{\hbar H_0^4}{24\pi^2c^3} \left( \frac{a_0}{a}\right)^4 
\left\{\ln \left[ \frac{2}{(a_0\tau) H_0} \left( \frac{a}{a_0} \right) \right]
 +\frac32\right\} +{\cal O}(a^{-5})
\ ,
\label{rhoq and pq:radiation:2}
\end{eqnarray}
where $a_0$ is the scale factor at $\eta_0$. 
Note that this result is invariant under a multiplicative rescaling of the 
scale factor, as required by coordinate invariance 
(this can be easily seen by recalling that $\tau$ mimics the comoving time
scale of the transition, such that $\tau_{\rm ph}=a_0\tau$ 
is the physical time scale of the transition). 
The classical radiation fluid energy density and pressure driving the expansion
of the background are
\begin{equation}
\rho = 3 p = \frac{3c^2 H_0^2}{8\pi G_N} \left( \frac{a_0}{a_{}} \right)^4 \ ,
\end{equation}
such that the quantum-to-classsical contribution scales at late times as
\begin{equation}
\frac{\rho_q}{\rho} \simeq 
 \frac{(\hbar H_0)^2}{3\pi (m_{\rm P}c^2)^2} 
    \left\{\ln\left[ \frac{2}{\tau_{\rm ph}H_0} 
              \left( \frac{a}{a_0} \right) \right] +\frac12\right\}
\ ,
\end{equation}
where the Planck mass is $m_{\rm P}=\sqrt{\hbar c/G_N}$.
The number of e-foldings after inflation $N=\ln(a/a_0)$
required for this ratio to become one is,
\begin{equation}
 N =  \frac{3\pi (m_{\rm P}c^2)^2}{(\hbar H_0)^2}  
                 + \ln\left( \frac{\tau_{\rm ph} H_0}{2{\rm e}^{1/2}} \right)
            \sim 10^{13} 
\,,
\label{N:qbr}
\end{equation}
where to get the numerical estimate we assumed that the scale of inflation
is $\hbar H_0 \sim 10^{13} \mathrm{GeV}$. Making the transition very fast
does not help much, since in a reasonable physical system
the transition time cannot be shorter than the Planck time, {\it i.e.}  
$\tau_{\rm ph}\geq t_{\rm P}$, for which  
$\ln(t_{\rm P}H_0)=\ln(\hbar H_0/m_{\rm P}c^2)\simeq -14\ll 10^{13}$.
On the other hand, pushing $H_0$ much closer to the Planck scale would
help, as it would drastically 
shorten the number of e-foldings $N$ at which 
the one-loop quantum contribution becomes significant.

One might wonder whether the logarithmic enhancement found 
in this section is of some spurious UV origin, since the late time
behavior depends on the regulating parameter $\tau$, that is
related to the details of the matching. 
The answer is: no. In fact, it is an IR effect.
Indeed, one can calculate the UV contribution to the energy density and
pressure, as was done in the previous section, and convince oneself 
that all of the contributions decay in time faster than the
background values. To summarize, the main result of this section is that,
based on the analysis presented in this work we can argue that, 
the sudden matching approximation suffices to 
unambiguously calculate the prefactor
of the leading term ($\propto \ln(a)/a^4$) in the one-loop energy density
and pressure at late times.
On the other hand, the subdominant 
contribution ($\propto 1/a^4$) will in general depend 
on the physical details of the transition, such as its rate and form.
In particular, the regulating parameter $\tau$, introduced to model
the time of transition, affects (logarithmically) 
the prefactor of that subleading contribution.

\subsection{Matching inflation to radiation to matter}
\label{Matching inflation to radiation to matter}

After discussing the subtleties of the sudden matching approximation and
resolving the UV issues (keeping in mind a more realistic physical picture), 
we are ready to present the main result of this paper. 
That is the late time behavior of the energy density and pressure of 
a massless, minimally coupled scalar field, that went through two sudden
transitions: one from inflation to radiation at $\eta_0$, and another
from a radiation era to a matter era at $\eta_1$. 
These transitions mimic the expansion history of the actual Universe 
and are illustrated as $\epsilon(t)$ 
in the left panel of figure~\ref{figure 1:epsilon}. At late times
($\eta\gg\eta_1$) the dominant one-loop contribution to the energy density
and pressure from these quantum fluctuations is
\begin{align}
\rho_q = \frac{3\mathcal{H}_0^4\mathcal{H}_1^2}{32\pi^2a^4\mathcal{H}^2} 
       +{\cal O}(\ln(a)/a^{4})
\ \ \ , \ \ \ \ \ 
p_q =       {\cal O}(\ln(a)/a^4)
\ ,
\label{matter era rhoq and pq}
\end{align}
where the full result is given in Appendix~D. It is instructive at this point 
to rewrite the energy density in terms of physical time $t$ and
the Hubble rate $H(t)=\dot{a}/a$, and to reinsert the units
\begin{equation}
\rho_q = \frac{3\hbar H_0^4}{32\pi^2c^3} \left( \frac{a_0}{a_1} \right)^4
      \left( \frac{a_1}{a_{}} \right)^3+{\cal O}(\ln(a)/a^4)
\,.
\label{final matter fluctuations:1}
\end{equation}
From here we see that our result does not depend on the choice of the scale 
factor value at some point in time, but only on the ratios of scale factors,
as it should be because of coordinate invariance. 
The classical non-relativistic matter
energy density required to drive the expansion during matter era is
\begin{equation}
\rho = \frac{3c^2 H_0^2}{8\pi G_N} \left( \frac{a_0}{a_1} \right)^4 
       \left( \frac{a_1}{a_{}} \right)^3 \ ,
\end{equation}
which allows us to express the ratio of the two energy densities 
at late times in terms of the scale of inflation,
\begin{equation}
\frac{\rho_q}{\rho} \simeq
 \frac{1}{4\pi}\frac{(\hbar H_0)^2}{(m_{\rm P}c^2)^2} 
  +{\cal O}(\ln(a)/a)
 \sim 10^{-13} 
\ ,
\label{final matter fluctuations:2}
\end{equation}
where $m_{\rm P}=\sqrt{\hbar c/G_N}$.
To get the last estimate we took for the Hubble energy scale at the end of 
inflation to be $\hbar H_0 \sim 10^{13}\mathrm{GeV}$. 
The contribution of $\rho_q$ in~(\ref{final matter fluctuations:2}) 
is unobservably small. 
Surprisingly, as it is shown in 
Appendix~C, the late time result~(\ref{final matter fluctuations:2})
is unaffected by the intermediate radiation era, see 
Eq.~(\ref{Appendix C: late time ratio}). 
It is also worth noting that at late times the
quantum energy density
$\rho_q$ scales as a non-relativistic matter, so that it constitutes 
an unmeasurably small contribution to dark matter (provided, of course, 
it clusters as dark matter, which ought to be checked).

\section{Discussion and conclusions}
\label{Discussion and conclusions}

 In this paper we study the quantum backreaction on the background 
geometry from the one-loop vacuum fluctuations of a massless minimally
coupled scalar field in de Sitter inflation, and the subsequent radiation
and matter era. We express our results in terms of the one-loop expectation 
value of the energy-momentum tensor with respect to the global Bunch-Davies
state in de Sitter space, defined in section~\ref{Scalar field on FLRW}.
 We observe that in the rest frame of the cosmological fluid 
the quantum energy momentum tensor is diagonal and isotropic, and hence can be 
represented in terms of equivalent one-loop contributions to 
the energy density and pressure, denoted as $\rho_q$ and $p_q$. 
 Our first important finding is a logarithmic enhancement 
of the one-loop energy density and pressure with respect to the cosmological
fluid in radiation 
era~(\ref{rhoq and pq:radiation}--\ref{rhoq and pq:radiation:2}).
Provided radiation era lasts long enough, and the scale of de Sitter inflation 
is high enough, this contribution will eventually dominate over the 
classical contribution. 
Our second most important finding is that in matter era 
scalar quantum fluctuations from de Sitter produce a tiny late time
contribution to the one-loop energy momentum 
tensor~(\ref{final matter fluctuations:1}--\ref{final matter fluctuations:2})
that scales as non-relativistic matter. 
Curiously, as shown in Appendix~C, the amplitude and scaling of this late 
time contribution is independent on the existence of an intermediate 
radiation era. Judged by the scaling alone,
this contribution then represents a small contribution to the dark matter of 
the Universe, provided it also clusters as dark matter. 
To answer the question of clustering, one would have to study 
the one-loop energy-momentum correlator, 
$\langle\Omega|\hat T_{\mu\nu}(x)\hat T_{\rho\sigma}(x^\prime)|\Omega\rangle$
in matter era, which is beyond the scope of this work.

An interesting extension of this work would be to consider the quantum 
backreaction from a slow roll inflation, in which during inflation 
$H(t)= H_0 a^{-\epsilon}$, where $\epsilon\ll 1$ and constant.
In this model one would generate a {\it red spectrum}
(defined as the equal time momentum space two-point correlator
multiplied by $k^3/(2\pi^2)$), 
that would enhance the quantum backreaction with respect to the flat spectrum
considered in this work. The important questions to answer
within this model would be whether the late time scaling
of the one-loop scalar quantum fluctuations in matter era would still scale 
as nonrelativistic matter and whether their amplitude would be enhanced
when compared with~(\ref{final matter fluctuations:2}).  
 
 Even though most of our calculations are done in a sudden matching
approximation, which generates unphysical boundary divergences 
due to the sudden matching procedure, in section~\ref{One sudden transition}
we discuss in detail how to regularize these by making use of a toy model
for smooth matching
in which ultraviolet particle production is suppressed as ${\rm e}^{-2\tau k}$,
where $\tau$ is a finite transition (conformal) time. 
We show by explicit calculation that this regulates all of the boundary 
divergences, and show that both in radiation and matter era 
the leading late time contribution is unaffected by $\tau$ 
(as long as $\tau$ is much shorter than the conformal Hubble time 
$\tau_{\cal H}=1/{\cal H}$ at the time of transition). 
This means that the late time results
from sudden matching are correct both in radiation and matter era. 

 While in this paper we discuss in detail the quantum backreaction
from massless minimally coupled scalar, in Appendix~E we construct 
a proof that the contribution from gravitons (when treated in the physical 
traceless, transverse gauge) equals two times the contribution from 
a massless minimally coupled scalar, such that our results 
for the MMCS can be directly translated to the results for the graviton
one-loop fluctuations. In particular, when 
Eq.~(\ref{final matter fluctuations:2}) is adapted to gravitons,
it yields in matter era, 
\begin{equation}
\frac{(\rho_q)_{\rm graviton}}{\rho} = 
2\frac{(\rho_q)_{\rm MCMS}}{\rho} \simeq
 \frac{1}{2\pi}\frac{(\hbar H_0)^2}{(m_{\rm P}c^2)^2} 
 \sim 10^{-13} 
\, ,
\label{final matter graviton fluctuations}
\end{equation}
still too small to be observable. 
When making the numerical estimate we have assumed that 
$\hbar H_0\sim 10^{13}~{\rm GeV}$. 
It is well known that 
inflationary fluctuations are dominated by the scalar cosmological 
perturbations. When compared with gravitons, scalars 
generate a spectrum enhanced by a factor $1/(16\epsilon)$,where 
$\epsilon=-\dot H/H^2$ is evaluated at the time when 
the modes in question cross the Hubble radius during inflation.
This can be used to make a crude estimate of the one-loop contribution 
to the quantum backreaction from scalar cosmological perturbations,
\begin{equation}
\frac{(\rho_q)_{\rm scalar\; perts.}}{\rho} 
\sim  \frac{1}{8\epsilon}\frac{(\rho_q)_{\rm MCMS}}{\rho} 
\simeq \frac{1}{32\pi\epsilon}\frac{(\hbar H_0)^2}{(m_{\rm P}c^2)^2} 
  \sim 10^{-12} 
\, ,
\label{final matter graviton fluctuations}
\end{equation}
which is still too small to be observable. 
When making the numerical estimate we have assumed that 
the relevant slow roll parameter $\epsilon\sim 10^{-2}$ and 
$\hbar H_0\sim 10^{13}~{\rm GeV}$. 
The late time matter era estimate~(\ref{final matter graviton fluctuations})
should not be confused with the estimate of 
the quantum backreaction during inflation 
from Refs.~\cite{Abramo:2001dc, Geshnizjani:2002wp}, 
in which case one can argue away 
any presence of one-loop quantum backreaction from cosmological perturbations
by choosing an appropriate observer. 
 We have thus presented a convincing case that the one-loop energy-momentum
tensor from quantum inflationary cosmological perturbations cannot account
for the dark energy of the late Universe, at least when 
the effect on the global Hubble rate is accounted for.
This result is to be contrasted with 
Refs.~\cite{KolbMatarreseRiotto, BarausseMatarreseRiotto}.
Even though our analysis is fully quantum, and those of 
Refs.~\cite{KolbMatarreseRiotto, BarausseMatarreseRiotto} classical
(that is they take no account of the quantum ultraviolet modes),
these results can be compared since the infrared cosmological perturbations 
are to a large extent classical.

Another possible extension of our work is to add
a non-minimal coupling to the scalar action~(\ref{scalar action}), 
$\Delta S_\xi = - \int d^Dx\sqrt{-g}(1/2)\xi {\cal R}\phi^2$, where 
${\cal R}$ denotes the Ricci scalar. 
During de Sitter inflation this term will generate a mass
for the scalar field. If $\xi<0$, the mass term will be negative, which 
will have as a consequence a red tilt in the spectrum
enhancing the effect of quantum fluctuations. Furthermore, while 
the presence of a nonminimal term will not change the mode functions 
for the global Bunch-Davies vacuum in radiation era, 
the mode functions of the global Bunch-Davies vacuum in matter era will change.
Hence it would be of interest to study the quantum backreaction
from these fluctuations at late times in matter era.

\section*{Acknowledgements}

DG would like to thank Igor Khavkine for a very usefull suggestion.

\section*{Appendix~A - Calculating the energy-momentum tensor
                       on smooth FLRW backgrounds}

Here we give an outline of the regularization and renormalization procedure 
needed to assign a finite physical value to integrals \eqref{intT00} 
and \eqref{intTij} representing the expectation value of
the one-loop energy-momentum tensor of the massless, minimally coupled scalar
field on smooth FLRW backgrounds. The first thing to do is to split
the integral into the IR part and the UV part by some constant cut-off
$\mu\gg\mathcal{H}$ on comoving momenta. The IR part of the integral 
can be evaluated immediately in $D=4$ (it might need an IR regulator, 
if the assumed state is not a physical one in the IR),
\begin{align}
\rho_q^{IR} ={}& \frac{1}{a^2} \langle\Omega |\hat{T}_{00}|\Omega\rangle_{IR} 
= \frac{1}{4\pi^2a^4} \int_0^\mu dk\, k^2 \left\{ |u'|^2 + \left[ k^2 
 + \mathcal{H}^2 \right] |u|^2 - \mathcal{H}\frac{\partial}{\partial\eta}|u|^2 
 \right\} 
\ , 
\end{align}
\begin{align}
p_q^{IR} ={}& \frac{1}{a^2} \langle\Omega |\hat{T}_{ij}|\Omega\rangle_{IR} 
= \frac{1}{4\pi^2a^4} \int_0^\mu dk\, k^2 \left\{ |u'|^2 
+ \left[ - \frac{k^2}{3} + \mathcal{H}^2 \right] |u|^2 
- \mathcal{H}\frac{\partial}{\partial\eta}|u|^2 \right\} \ .
\end{align}
The UV part of the integral exhibits the usual quartic, quadratic
and logarithmic divergences, therefore we first need to regularize it. 
We employ dimensional regularization, where power-law divergences 
are automatically subtracted, leaving only the logarithmic one. 
The UV contributions are of the form, 
\begin{align}
&\langle\Omega |\hat{T}_{00}|\Omega\rangle_{UV} =  \frac{a^{2-D}}{(4\pi)^{\frac{D-1}{2}}\Gamma\left( \frac{D-1}{2} \right)} \int_{\mu}^\infty dk \, k^{D-2} \Bigg{\{} |u'|^2 + \left[k^2 + \frac{(D\!-\!2)^2}{4}\mathcal{H}^2\right]|u|^2 - \frac{D\!-\!2}{2}\mathcal{H}\frac{\partial}{\partial\eta}|u|^2 \Bigg{\}} \ , \label{UV00} \\
& \langle\Omega |\hat{T}_{ij}|\Omega\rangle_{UV} = \nonumber \\
& \ \ \ \ = \frac{\delta_{ij} a^{2-D}}{(4\pi)^{\frac{D-1}{2}}\Gamma\left( \frac{D-1}{2} \right)} \int_{\mu}^{\infty} dk\, k^{D-2} \Bigg{\{} |u'|^2 + \left[- \frac{D\!-\!3}{D\!-\!1}k^2 + \frac{(D\!-\!2)^2}{4}\mathcal{H}^2 \right] |u|^2 - \frac{D\!-\!2}{2}\mathcal{H}\frac{\partial}{\partial\eta}|u|^2 \Bigg{\}} \ . \label{UVij}
\end{align}
The two integrals above need to be calculated to order $\mathcal{O}(\mu^{-1})$, since in the final expression we take the limit $\mu\rightarrow\infty$. In order to do that we first find the asymptotic expansion of the mode functions $u(k,\eta)$ as a power series in $1/k$, for $k\ge \mu \gg \mathcal{H}$. That can be obtained from the WKB approximation of the equation of 
motion~\eqref{mode function EOM} for the scalar mode function. A shortcut way of doing it is writing down the asymptotic expansion for the positive-frequency mode function in the form
\begin{equation}
u(k,\eta) = \frac{e^{-ik\eta}}{\sqrt{2k}} \left[ 1 + \frac{iF_1(\eta)}{k} + \frac{F_2(\eta)}{k^2} + \frac{iF_3(\eta)}{k^3} + \frac{F_4(\eta)}{k^4} \right] + \mathcal{O}(k^{-5}) 
\ ,
\label{symb asympt}
\end{equation}
where all the $F_i(\eta)$ functions can now be obtained by solving the equation
of motion and satisfying the Wronskian normalization order by order in
$1/k$ up to $\mathcal{O}(k^{-5})$. The equation of motion, 
up to the desired order, gives four coupled differential equations,
\begin{align}
& F_1'(\eta) = -\frac{1}{2}f(\eta) \ , \label{eomF1} \\
& F_2'(\eta) = \frac{1}{2} \Big{[} F_1''(\eta) + f F_1(\eta) \Big{]} \ , \label{eomF2} \\
& F_3'(\eta) = -\frac{1}{2} \Big{[} F_2''(\eta) + f(\eta) F_2(\eta) \Big{]} \ , \label{eomF3} \\
& F_4'(\eta) = \frac{1}{2} \Big{[} F_3''(\eta) + f(\eta) F_3(\eta) \Big{]} \ , \label{eomF4}
\end{align}
where
\begin{equation}
f(\eta) = -\frac{(D-2)}{4}\Big{[} 2\mathcal{H}' + (D-2)\mathcal{H}^2 \Big{]} \ ,
\end{equation}
and the Wronskian condition gives two equations
\begin{align}
& 2F_2 + F_1^2 - F_1' = 0 \ , \label{wr1} \\
& 2 F_4 + 2F_3F_1 + F_2^2 - F_3' +F_2'F_1 - F_1'F_2 = 0 \ . \label{wr2}
\end{align}
Integrating \eqref{eomF1}--\eqref{eomF4} and imposing conditions \eqref{wr1}, 
\eqref{wr2} is straightforward. By assuming convenient initial conditions, namely  
that all the $F_n(\eta)$ functions for odd $n$ are zero at some arbitrary
point in time $\eta_0$ (which does not affect physical quantities),~\footnote{
This procedure amounts to choosing a time independent phase for the mode 
function. Here a convenient choice is made, but of course it
is not unique. In sections~\ref{Matching inflation to radiation to matter}
and~\ref{Matching inflation to radiation} 
a different choice for the phase is made. 
Of course, this phase does not affect physical quantities. 
} 
the functions in \eqref{symb asympt} turn out to be
\begin{align}
& F_1(\eta) = -\frac{1}{2} \int_{\eta_0}^{\eta}d\tau f(\tau) \ , 
\label{F1}\\
& F_2(\eta) = -\frac{f(\eta)}{4} 
- \frac{1}{8} \left[ \int_{\eta_0}^{\eta}d\tau f(\tau) \right]^2 \ , 
\label{F2}\\
& F_3(\eta) = \frac{1}{8} \left( f'(\eta) - f'(\eta_0) \right) 
+ \frac{f(\eta)}{8}\int_{\eta_0}^{\eta}d\tau f(\tau) 
+ \frac{1}{48} \left[ \int_{\eta_0}^{\eta}d\tau f(\tau) \right]^3 
+ \frac{1}{8} \int_{\eta_0}^{\eta}d\tau f^2(\tau) \ , 
\label{F3}\\
& F_4(\eta) = \frac{f''(\eta)}{16} + \frac{5 f^2(\eta)}{32} 
+ \frac{1}{16} \left( f'(\eta) - f'(\eta_0) \right) 
\int_{\eta_0}^{\eta}d\tau f(\tau) + \frac{f(\eta)}{32} 
\left[ \int_{\eta_0}^{\eta}d\tau f(\tau) \right]^2  \nonumber \\
& \ \ \ \ \ \ \ \ \ + \frac{1}{384} 
\left[ \int_{\eta_0}^{\eta}d\tau f(\tau) \right]^4 
+ \frac{1}{16} \left[ \int_{\eta_0}^{\eta}d\tau f(\tau) \right] 
\left[ \int_{\eta_0}^{\eta}d\tau f^2(\tau) \right] \ .
\label{F4}
\end{align}
The higher order functions soon become unprintably complicated.

Now we can easily evaluate \eqref{UV00} and \eqref{UVij}, expand them around $D=4$, and discard the terms $\mathcal{O}(D-4)$ and $\mathcal{O}(\mu^{-1})$ right away,
\begin{align}
\langle\Omega |T_{00}|\Omega\rangle_{UV} ={}&  \frac{1}{16\pi^2a^2} \Bigg{\{} -\mu^4 - \mathcal{H}^2\mu^2 - \frac{1}{2}\Big{[}-2\mathcal{H}''\mathcal{H} + (\mathcal{H}')^2 + 3\mathcal{H}^4\Big{]} \frac{\mu^{D-4}}{D-4} + \nonumber  \\
&  + \frac{1}{2}(\ln a - c) \Big{[}-2\mathcal{H}''\mathcal{H} + (\mathcal{H}')^2 + 3\mathcal{H}^4 \Big{]} - \frac{3}{2}\mathcal{H}^4 \Bigg{\}} + \mathcal{O}(\mu^{-2}) \ , \label{T00UV} \\
\langle\Omega |T_{ij}|\Omega\rangle_{UV} ={}& \frac{\delta_{ij}}{48\pi^2a^2} \Bigg{\{} - \mu^4 + (2\mathcal{H}' - \mathcal{H}^2)\mu^2  - \nonumber \\
& - \frac{1}{2} \Big{[} 2\mathcal{H}'''-2\mathcal{H}''\mathcal{H}+(\mathcal{H}')^2 -12 \mathcal{H}'\mathcal{H}^2 + 3\mathcal{H}^4 \Big{]} \frac{\mu^{D-4}}{D-4} + \nonumber \\
& + \frac{1}{2} (\ln a - c) \Big{[} 2\mathcal{H}''' - 2\mathcal{H}''\mathcal{H} + (\mathcal{H}')^2 - 12\mathcal{H}'\mathcal{H}^2 + 3\mathcal{H}^4 \Big{]} + \nonumber \\
&+ \frac{1}{6} \Big{[} 2\mathcal{H}''' - 2\mathcal{H}''\mathcal{H} + (\mathcal{H}')^2 +24\mathcal{H}'\mathcal{H}^2 - 6\mathcal{H}^4 \Big{]} \Bigg{\}} + \mathcal{O}(\mu^{-2}) \ , \label{TijUV} 
\end{align}
where $c=\frac{1}{2}(\gamma_E-\ln\pi)$.

The counterterm in the action needed to absorb the logarithmic divergence is
\begin{equation}
S_{ct} = \alpha S_1 = \alpha \int d^Dx \, \sqrt{-g} \, R^2 \ ,
\end{equation}
where
\begin{equation}
\alpha = \frac{1}{1152\pi^2} \left( \frac{\mu^{D-4}}{D-4} + \alpha_f \right)
 \ ,
\label{counterterm:R2}
\end{equation}
and $\alpha_f$ is a free finite constant
(that can depend on $\mu$ logarithmically), 
in principle determined by measurements. 
The contribution to the energy-momentum tensor coming from this counterterm is
\begin{equation} \label{alpha(1)H}
\alpha \times^{(1)}{H}{_{\mu\nu}} 
 = \alpha \frac{-2}{\sqrt{-g}} \frac{\delta S_1}{\delta g^{\mu\nu}} 
 = \alpha (4 \nabla_{\mu} \nabla_{\nu} R - 4g_{\mu\nu} \square R 
  + g_{\mu\nu}R^2 - 4R_{\mu\nu} R) 
\ .
\end{equation}
On FLRW backgrounds ${}^{(1)}{H}{_{\mu\nu}}$ is diagonal, 
and, when expanded around $D=4$, the contribution \eqref{alpha(1)H} reads
\begin{align}
\alpha\times{}^{(1)}{H}{_{00}} 
={}& \frac{1}{32\pi^2a^2} \Big{[} -2\mathcal{H}''\mathcal{H} 
+ (\mathcal{H}')^2 + 3\mathcal{H}^4 \Big{]} \frac{\mu^{D-4}}{D-4} 
\nonumber \\
&+ \frac{1}{96\pi^2a^2}\Big{[} -4\mathcal{H}''\mathcal{H} 
+ 2(\mathcal{H}')^2 - 6\mathcal{H}'\mathcal{H}^2 + 9\mathcal{H}^4 \Big{]}  
\nonumber \\
& + \frac{\alpha_f}{32\pi^2a^2} \Big{[} -2\mathcal{H}''\mathcal{H} 
 + (\mathcal{H}')^2 + 3\mathcal{H}^4 \Big{]}  \ , \label{1H00} \\
\alpha\times{}^{(1)}{H}{_{ij}} 
={}& \frac{\delta_{ij}}{96\pi^2a^2} \Big{[} 2\mathcal{H}''' 
  - 2\mathcal{H}''\mathcal{H} + (\mathcal{H}')^2 -12\mathcal{H}'\mathcal{H}^2
  + 3\mathcal{H}^4 \Big{]} \frac{\mu^{D-4}}{D-4} 
+ \nonumber \\
&+ \frac{\delta_{ij}}{288\pi^2a^2} \Big{[} 2\mathcal{H}''' 
+10\mathcal{H}''\mathcal{H} + 10(\mathcal{H}')^2 
- 30\mathcal{H}'\mathcal{H}^2 - 3\mathcal{H}^4 \Big{]} 
\nonumber \\
&+  \frac{\delta_{ij}\alpha_f}{96\pi^2a^2} \Big{[} 2\mathcal{H}''' 
- 2\mathcal{H}''\mathcal{H} + (\mathcal{H}')^2 -12\mathcal{H}'\mathcal{H}^2 
+ 3\mathcal{H}^4 \Big{]} \ . \label{1Hij}
\end{align}
Now we see that the terms that are divergent for $D\rightarrow4$ 
in expressions \eqref{1H00} and \eqref{1Hij} above exactly cancel 
the divergent terms from \eqref{T00UV} and \eqref{TijUV}.

 There is one more contribution in the UV part of the energy momentum tensor, 
the so-called 
{\it conformal anomaly} (CA)~\cite{Capper:1974ic,Duff:1977ay,Birrell:1982ix}.
Although that contribution is strictly
speaking not necessary to renormalize the energy-momentum tensor 
on FLRW background, it is necessary for the renormalization 
on more general backgrounds and we include it here. 
Its contribution on FLRW background is
\begin{align}
T^{CA}_{00} ={}& \frac{1}{2880\pi^2a^2} \Big{[} 
   2\mathcal{H}''\mathcal{H} - (\mathcal{H}')^2 \Big{]} 
  + \frac{3\alpha_{CA}}{a^2} \Big{[} 2\mathcal{H}''\mathcal{H} 
- (\mathcal{H}')^2 - 3\mathcal{H}^4 \Big{]} 
\ , \\
T^{CA}_{ij} ={}& -\frac{\delta_{ij}}{8640\pi^2a^2} \Big{[} 2\mathcal{H}''' 
 - 2\mathcal{H}''\mathcal{H} + (\mathcal{H}')^2 \Big{]} 
\nonumber\\
{}&+ \frac{\delta_{ij}\alpha_{CA}}{a^2} \Big{[} -2\mathcal{H}''' 
+ 2\mathcal{H}''\mathcal{H} - (\mathcal{H}')^2 + 12\mathcal{H}'\mathcal{H}^2
 -3\mathcal{H}^4 \Big{]} \ ,
\end{align} 
where $\alpha_{CA}$ is a free constant. 

Finally, adding all these three contributions together, 
the renormalized UV part of the one-loop expectation value of
the energy-momentum tensor is
\begin{align}
T_{00}^{UV} ={}& \lim_{D\rightarrow4} 
\Big{(} \langle\Omega | T_{00}|\Omega\rangle^{UV} 
  + \alpha \times{}^{(1)}{H}{_{00}} + T^{CA}_{00} \Big{)} 
\nonumber \\
={}&  \frac{1}{16\pi^2a^2} \Bigg{\{}- \mu^4 - \mathcal{H}^2\mu^2 
+ \frac{1}{2} (\ln a + \tilde\alpha_f) \Big{[} -2\mathcal{H}''\mathcal{H} 
+ (\mathcal{H}')^2 + 3\mathcal{H}^4 \Big{]} 
\nonumber \\
&+ \frac{1}{3} \Big{[} -2\mathcal{H}''\mathcal{H} +(\mathcal{H}')^2 
-3\mathcal{H}'\mathcal{H}^2 \Big{]} 
+ \frac{1}{180} \Big{[} 2\mathcal{H}''\mathcal{H} - (\mathcal{H}')^2 \Big{]} 
\ , 
\end{align}
\begin{align}
T_{ij}^{UV} ={}& \lim_{D\rightarrow 4} \Big{(}  
   \langle\Omega | T_{ij}|\Omega\rangle^{UV} + \alpha \times{}^{(1)}{H}{_{ij}} 
 + T^{CA}_{ij} \Big{)}
\\
={}& \frac{\delta_{ij}}{48\pi^2a^2} \Bigg{\{} - \mu^4 
+ (2\mathcal{H}'-\mathcal{H}^2)\mu^2 
 + \frac{1}{2} (\ln a + \tilde\alpha_f) \Big{[} 2\mathcal{H}''' 
           - 2\mathcal{H}''\mathcal{H} + (\mathcal{H}')^2 
           -12\mathcal{H}'\mathcal{H}^2 + 3\mathcal{H}^4 \Big{]} 
\nonumber \\
&+ \frac{1}{6} \Big{[} 4 \mathcal{H}''' + 8\mathcal{H}''\mathcal{H} 
+ 11(\mathcal{H}')^2 - 6\mathcal{H}'\mathcal{H}^2 - 9\mathcal{H}^4 \Big{]} 
- \frac{1}{180} \Big{[} 2\mathcal{H}''' - 2\mathcal{H}''\mathcal{H} 
+ (\mathcal{H}')^2 \Big{]} \Bigg{\}} \ ,
 \nonumber
\end{align}
where $c$ and $\alpha_{CA}$ have been absorbed into $\tilde\alpha_f$. 
The tensor defined above satisfies the conservation equation by itself, 
which is to be expected since we have chosen a time-independent cut-off
to separate IR from UV. Specializing the result to FLRW backgrounds with
constant $\epsilon$, the UV parts of the energy density and pressure read
\begin{align}
\rho_q^{UV} ={}& \frac{1}{16\pi^2a^4} \bigg\{\!-\!\mu^4\!-\!\mathcal{H}^2\mu^2 
+ \frac{3\mathcal{H}^4}{2} \epsilon(2\!-\!\epsilon)[\ln(a)\!+\!\tilde\alpha_f] 
- \mathcal{H}^4(1\!-\!\epsilon)(2\!-\!\epsilon) 
  + \frac{\mathcal{H}^4}{60}(1\!-\!\epsilon)^2 \bigg{\}} 
\ , \\
p_q^{UV} ={}& \frac{\delta_{ij}}{48\pi^2a^4} 
 \bigg\{ \!-\! \mu^4 + (1\!-\!2\epsilon)\mathcal{H}^2\mu^2 
   - \frac{3\mathcal{H}^4}{2} \epsilon(2\!-\!\epsilon)(3\!-\!4\epsilon)
            [\ln(a)\!+\!\tilde\alpha_f]
\nonumber \\
&\hskip 1.5cm
+ \frac{\mathcal{H}^4}{2} (2\!-\!\epsilon)(6\!-\!17\epsilon\!+\!8\epsilon^2) 
  - \frac{\mathcal{H}^4}{60}(1\!-\!\epsilon)^2(3\!-\!4\epsilon) \bigg{\}}
\,.
\end{align}
The full answer is obtained by adding the UV and  IR contributions,
\begin{align}
\rho_q ={}& \lim_{\mu\rightarrow\infty} \Big{(} \rho^{IR}_q + \rho_q^{UV} \Big{)} \ , \\
p_q ={}& \lim_{\mu\rightarrow\infty} \Big{(} p_q^{IR} + p_q^{UV} \Big{)} \ .
\end{align}

In the end, we comment on the statement made in the beginning of 
section~\ref{Energy-momentum on FLRW with evolving epsilon}. 
There it was stated that it makes no difference for the physical result
whether one works with a $D$-dependent or $D$-independent $\epsilon$ parameter.
Let us now examine expressions \eqref{T00UV}, \eqref{TijUV} and 
\eqref{1H00}, \eqref{1Hij}. The factors that multiply the $1/(D-4)$ divergence
are the same, just with the opposite sign, such that they cancel each other when
the two contributions are added together. These factors are still $D$ 
dependent, because they consist of geometric quantities of $D$-dimensional 
FLRW spacetime. But in practice it is not necessary to expand them around
$D=4$, since the finite terms this would generally cancel exactly in the final
result. 

To make this important point clearer, let us consider the divergent terms in
\eqref{T00UV}, \eqref{TijUV}, and \eqref{1H00}, \eqref{1Hij}
on a constant $\epsilon$ background. The divergent terms are
\begin{align}
& \Big{[} \langle\Omega | T_{00}|\Omega\rangle_{UV} \Big{]}_{\rm div} 
=-\frac{3\mathcal{H}^4}{32\pi^2a^2}
     \epsilon(2\!-\!\epsilon)\frac{\mu^{D-4}}{D\!-\!4}
= - \Big{[} \alpha \times{}^{(1)}\!{H}{_{00}} \Big{]}_{\rm div} 
\ , \\
& \Big{[} \langle\Omega | T_{ij}|\Omega\rangle_{UV} \Big{]}_{\rm div} 
= - \frac{\mathcal{H}^4}{32\pi^2a^2} \epsilon(2\!-\!\epsilon)(3\!-\!4\epsilon) 
          \frac{\mu^{D-4}}{D\!-\!4} 
= - \Big{[} \alpha \times{}^{(1)}\!{H}{_{ij}} \Big{]}_{\rm div} 
\ .
\end{align}
Now, if we assume $\epsilon$ to be $D$-dependent, 
its expansion around $D=4$ being 
$\epsilon = \epsilon_4 + \delta\epsilon (D-4)$, it is clear 
that in the final answer there will be no terms that depend 
on $\delta\epsilon$, since they cancel when we add the two contributions above.

\section*{Appendix B - Integrals}

Here we list all the non-trivial integrals used in
sections~\ref{Energy-momentum tensor on FLRW with constant epsilon}
and~\ref{Energy-momentum on FLRW with evolving epsilon}. 
Even though, there is no IR divergence in the final answers for 
the energy density and pressure, individual integrals are IR divergent, 
and we regularize them with an IR cut-off $k_0$. 
We discard the terms of order $\mathcal{O}(k_0)$, 
and keep the regulating parameter $\tau$ only where it is necessary
to regulate the result for 
$\Delta
\rightarrow 0$, 
and discard the fully regular $\mathcal{O}(\tau)$ terms. 
The relevant integrals are of the form,
\begin{align}
\mathcal{I}_{2n} = \int_{k_0}^{\infty} dk \, k^{-2n} 
\sin(k\Delta) {\rm e}^{-\tau k} 
\ , \ \ \ 
\mathcal{I}_{2n-1} = \int_{k_0}^{\infty} dk\, k^{1-2n} \cos(k\Delta) 
{\rm e}^{-\tau k} 
\ , \ \ \ (n=0,1,\dots) \ ,
\end{align}
and their values are
\begin{align}
& \mathcal{I}_{-1} = -\frac{\Delta^2 - \tau^2}{(\Delta^2 + \tau^2)^2} \ , \nonumber \\
&  \mathcal{I}_0 = \frac{\Delta}{\Delta^2+\tau^2} \ , \nonumber \\
& \mathcal{I}_1 = - \gamma_E - \frac{1}{2} \ln\big{[} k_0^2(\Delta^2+\tau^2) \big{]} \ , \\
& \mathcal{I}_2 = \Delta \left[ 1 - \gamma_E - \frac{1}{2}\ln\big{[} k_0^2(\Delta^2+\tau^2) \big{]} \right] \  , \nonumber \\
& \mathcal{I}_3 = \frac{1}{2k_0^2} + \frac{\Delta^2}{2} \left[ -\frac{3}{2} + \gamma_E + \frac{1}{2}\ln\big{[} k_0^2(\Delta^2+\tau^2) \big{]} \right] \ , \\
& \mathcal{I}_4 = \frac{\Delta}{2k_0^2} + \frac{\Delta^3}{6} \left[ - \frac{11}{36} + \gamma_E + \frac{1}{2}\ln\big{[} k_0^2(\Delta^2+\tau^2) \big{]} \right] \ , \\
& \mathcal{I}_5 = \frac{1}{4k_0^4} - \frac{\Delta^2}{4k_0^2} + \frac{\Delta^4}{24} \left[ \frac{25}{12} - \gamma_E - \frac{1}{2} \ln\big{[} k_0^2(\Delta^2+\tau^2) \big{]} \right] \ , \\
& \mathcal{I}_6 = \frac{\Delta}{4k_0^4} - \frac{\Delta^3}{12k_0^2} + \frac{\Delta^5}{120} \left[ \frac{137}{60} - \gamma_E - \frac{1}{2}\ln\big{[} k_0^2(\Delta^2+\tau^2) \big{]} \right] \ , \\
& \mathcal{I}_7 = \frac{1}{6k_0^6} - \frac{\Delta^2}{8k_0^4} + \frac{\Delta^4}{48k_0^2} + \frac{\Delta^6}{720} \left[ - \frac{49}{20} + \gamma_E + \frac{1}{2}\ln\big{[} k_0^2(\Delta^2+\tau^2) \big{]} \right] \ , \\
& \mathcal{I}_8 = \frac{\Delta}{6 k_0^6} - \frac{\Delta^3}{24k_0^4} + \frac{\Delta^5}{240k_0^2} + \frac{\Delta^7}{5040} \left[ - \frac{363}{140} + \gamma_E + \frac{1}{2} \ln\big{[} k_0^2(\Delta^2+\tau^2) \big{]} \right] \ , \\
& \mathcal{I}_9 = \frac{1}{8k_0^8} - \frac{\Delta^2}{12k_0^6} + \frac{\Delta^4}{96k_0^4} - \frac{\Delta^6}{1440k_0^2} + \frac{\Delta^8}{40320} \left[ \frac{761}{280} - \gamma_E - \frac{1}{2} \ln\big{[} k_0^2(\Delta^2+\tau^2) \big{]} \right] \ .
\end{align}

\section*{Appendix C - Inflation to matter}

Here we present the result for the one-loop energy density and pressure
of quantum fluctuations (represented by a massless, minimally coupled scalar
field) living on a FLRW background that went through a sudden transition
from an inflationary period to a matter-dominated period at some time
$\eta_0$ ($\mathcal{H}(\eta_0)=\mathcal{H}_0$). The calculation is performed
in the same manner as the one presented in section~\ref{Radiation}.
Since we are interested in the late time behavior, 
the presented result is regulated only for times well after the matching,
\begin{align}
\rho_q ={}& \frac{3}{128\pi^2a^4} \Bigg{\{} \frac{2\mathcal{H}_0^3\mathcal{H}^2}{\mathcal{H}_0-\mathcal{H}} + 3\mathcal{H}_0^4 \ln \left[ \frac{4}{\tau} \frac{(\mathcal{H}_0-\mathcal{H})}{\mathcal{H}_0\mathcal{H}} \right] + 3\mathcal{H}^4 \ln \left[ 4\mu_f a \frac{(\mathcal{H}_0-\mathcal{H})}{\mathcal{H}_0\mathcal{H}} \right] + \nonumber \\
& + \left[ 3\gamma_E - \frac{913}{180} \right] \mathcal{H}^4 + 3 \mathcal{H}_0\mathcal{H}^3 - \frac{1}{2}\mathcal{H}_0^2\mathcal{H}^2 + 3 \mathcal{H}_0^3\mathcal{H} - \frac{15}{4}\mathcal{H}_0^4 + 4 \frac{\mathcal{H}_0^6}{\mathcal{H}^2} \Bigg{\}} \ , \\
p_q ={}& \frac{3}{128\pi^2a^4} \Bigg{\{} \frac{\mathcal{H}_0^4\mathcal{H}^2}{3(\mathcal{H}_0-\mathcal{H})^2} + \mathcal{H}_0^4 \ln\left[ \frac{4}{\tau} \frac{(\mathcal{H}_0-\mathcal{H})}{\mathcal{H}_0\mathcal{H}} \right] + 3\mathcal{H}^4 \ln \left[ 4\mu_f a \frac{(\mathcal{H}_0-\mathcal{H})}{\mathcal{H}_0\mathcal{H}} \right] + \nonumber \\
& + \left[ 3\gamma_E - \frac{1093}{180} \right] \mathcal{H}^4 + 3\mathcal{H}_0\mathcal{H}^3 + \frac{1}{6}\mathcal{H}_0^2\mathcal{H}^2 + \mathcal{H}_0^3\mathcal{H} - \frac{7}{4}\mathcal{H}_0^4 \Bigg{\}} \ .
\end{align}
We want to describe well only the late time behavior. 
Since $\mathcal{H}\sim a^{-1/2}$, the dominant late time behavior
of the expressions above is
\begin{equation}
\rho_q = \frac{3\mathcal{H}_0^6}{32\pi^2a^4\mathcal{H}^2} 
       + {\cal O}(\ln(a)/a^{4}) 
\ \ , \ \ \ \ \ 
p_q = {\cal O}(\ln(a)/a^{4}) \ ,
\label{Appendix C: rhoq pq}
\end{equation}
for which the late-time equation of state parameter vanishes,
\begin{equation}
w_q = 0 = w \ .
\end{equation}
Note that this is the same late-time behavior as is obtained 
in~(\ref{matter era rhoq and pq}) of 
subsection~\ref{Matching inflation to radiation to matter}.
Written in terms of physical Hubble rate and with dimensionfull constants
reinserted, the result~(\ref{Appendix C: rhoq pq}) becomes
\begin{equation}
\rho_q = \frac{3\hbar H_0^4}{32 \pi^2 c^3} 
\left( \frac{a_0}{a_{}} \right)^3  + {\cal O}(\ln(a)/a^{4})
\ \ , \ \ \ \ \ 
p_q ={\cal O}(\ln(a)/a^{4}) \ .
\end{equation}
The late time ratio of the energy density of one-loop quantum fluctuations 
to the energy density of classical matter fluid driving the expansion after
the transition is
\begin{equation}
\frac{\rho_q}{\rho} = \frac{G_N\hbar H_0^2}{4\pi c^5} + {\cal O}(\ln(a)/a) 
   = \frac{1}{4\pi}\left(\frac{\hbar H_0}{m_{\rm P}c^2}\right)^2
      + {\cal O}(\ln(a)/a) 
\ .
\label{Appendix C: late time ratio}
\end{equation}
Note that this is the same ratio as obtained in 
Eq.~(\ref{final matter fluctuations:2}) 
in subsection~\ref{Matching inflation to radiation to matter},
where we had an intermediate period of radiation-dominated expansion. 
Apparently, when compared with the background fluid density, 
this intermediate radiation period (no matter how long it lasts) neither 
amplifies nor damps the relative contribution
of the late time one-loop vacuum fluctuations.

\section*{Appendix D - The complete result for the inflation-radiation-matter
                       transition: matter era}

In this Appendix we give the full results for the one-loop energy density
and pressure of the scalar field on the FLRW background that has gone through
two sudden transitions, from a global Bunch-Davies state in de Sitter inflation
to a radiation era at $\eta_0$ ($\mathcal{H}(\eta_0)=\mathcal{H}_0$), and from
radiation to a matter era at $\eta_1$ 
($\mathcal{H}(\eta_1)=\mathcal{H}_1$). The calculation is analogous to the
one performed 
in section~\ref{Energy-momentum tensor on FLRW with constant epsilon}, 
and we have used the integrals from appendix~B. 
Since we are interested in the late time behavior, 
the result presented here is regulated for times well after 
the second matching, meaning we have set the UV regulator 
$\tau$ (which mimics the transition time on both transitions) 
to zero everywhere where it can be done without a consequence for 
late times. In the expressions below we use 
$\zeta={\mathcal{H}}/{\mathcal{H}_1}$ 
and $\zeta_0={\mathcal{H}_1}/{\mathcal{H}_0}$, such that 
$\zeta_0\ll 1 $ and at late times 
$\zeta\rightarrow 0$ (when $\eta\rightarrow\infty$).

The results for $\rho_q$ and $p_q$ are:
\begin{align}
 \frac{4\pi^2 a^4}{\mathcal{H}_0^4} \rho_q
&= - \frac{ \zeta_0^4 \zeta^2}{16(1-\zeta)} 
+ \frac{\zeta_0^2 \zeta^2}{2(2-\zeta-\zeta_0\zeta)}
 + \frac{1}{10080 \zeta^2} \Big{[} 3239 + 506 \zeta_0 + 35\zeta_0^2 \Big{]} 
\nonumber \\
& - \frac{(1-\zeta_0)}{5040\zeta} \Big{[} 253 \!+\! 35\zeta_0 \Big{]} + \frac{1}{80640} \Big{[} \!-\!47153 \!+\! 14636 \zeta_0 \!+\! 13758 \zeta_0^2 \!-\!7372 \zeta_0^3 \!-\! 2219 \zeta_0^4 \Big{]} 
\nonumber \\
& + \frac{\zeta}{20160} \Big{[} 1689 + 1204 \zeta_0 + 74 \zeta_0^2 + 2444 \zeta_0^3 + 259 \zeta_0^4 \Big{]} 
 \\
& + \frac{\zeta^2}{161280} \Big{[} 2011 + 4254 \zeta_0 - 43459 \zeta_0^2 + 31156 \zeta_0^3 - 4195 \zeta_0^4 + 2862 \zeta_0^5 - 189 \zeta_0^6 \Big{]} 
\nonumber \\
& + \frac{3\zeta^3}{8960} \Big{[} -105 + 310 \zeta_0 - 391 \zeta_0^2 + 84 \zeta_0^3 + 1041 \zeta_0^4 - 106 \zeta_0^5 + 7\zeta_0^6 \Big{]} + \nonumber \\
& + \frac{\zeta^4}{53760} \Big{[} 135 \!+\! 1890 \zeta_0 \!-\! 1845 \zeta_0^2 \!-\! 13500 \zeta_0^3 \!+\! 5(3024\gamma_E\!-\!2627)\zeta_0^4 \!+\! 954 \zeta_0^5\!-\! 63 \zeta_0^6 \Big{]}
\nonumber \\
& + \frac{9\zeta_0^4\zeta^4}{32} \ln \left[ 2a \left( \frac{\mu_f}{\mathcal{H}_0}\right) \frac{1-\zeta_0}{\zeta_0}  \right] + \left( \frac{1}{2} + \frac{\zeta_0^4}{32} \right) \ln \left[ \frac{1-\zeta_0}{\tau\mathcal{H}_1} \right] 
\nonumber \\
& + \left\{ - \frac{1}{36\zeta^4} - \frac{8}{45\zeta} + \frac{1}{2} + \frac{\zeta_0^4}{32} - \frac{4\zeta^2}{9} + \zeta^4\left[ \frac{9\zeta_0^4}{32} + \frac{3}{20} \right] \right\} \ln\left[ \frac{2(1-\zeta)}{(1-\zeta_0)\zeta} \right]
\nonumber \\
&+ \Bigg{\{} \frac{1}{72\zeta^4}\!+\! \frac{2}{35\zeta^3}\!+\! \frac{ \zeta_0^3 -11}{90\zeta}\!+\! \frac{5\!+\! 4\zeta_0^3\!-\! \zeta_0^4}{64}\!+\! \zeta^2 \left[\frac{697}{2880} \!-\!\frac{11\zeta_0^3}{144}\!+\! \frac{9\zeta_0^4}{64}\!-\! \frac{3\zeta_0^5}{80}\!+\! \frac{\zeta_0^6}{288}  \right] 
\nonumber \\
& + \zeta^4 \left[ - \frac{4449}{71680}\!-\! \frac{141\zeta_0^3}{1280}\!+\! \frac{45\zeta_0^4}{1024}\!+\! \frac{33\zeta_0^5}{640}\!-\! \frac{9\zeta_0^6}{256}\!+\! \frac{9\zeta_0^7}{1792} \!-\! \frac{3\zeta_0^8}{10240} \right] \Bigg{\}} \ln\left[ 1 \!+\! \frac{2(1\!-\!\zeta)}{(1\!-\!\zeta_0)\zeta}  \right] 
\nonumber \\
&+ \Bigg{\{} \frac{1}{72\zeta^4} \!-\! \frac{2}{35\zeta^3} \!+\! \frac{27\!-\!\zeta_0^3}{90\zeta} \!+\! \frac{-27 \!+\! 4\zeta_0^3\!-\!\zeta_0^4}{64} \!+\! \zeta^2\left[ \frac{81}{320} \!-\! \frac{3\zeta_0^3}{16}\!+\! \frac{9\zeta_0^4}{64}\!-\! \frac{3\zeta_0^5}{80}\!+\! \frac{\zeta_0^6}{288} \right]
\nonumber \\
&  + \zeta^4\! \left[-\frac{6561}{71680} \!+\! \frac{243 \zeta_0^3}{1280}\!-\! \frac{243\zeta_0^4}{1024}\!+\! \frac{81\zeta_0^5}{640}\!-\! \frac{9\zeta_0^6}{256} \!+\! \frac{9\zeta_0^7}{1792}\!-\! \frac{3\zeta_0^8}{10240} \right] \Bigg{\}} \!\ln\left[ \left|1\!-\! \frac{2(1\!-\!\zeta)}{(1\!-\!\zeta_0)\zeta} \right|  \right],
\nonumber
\end{align}
\begin{align}
 \frac{4\pi^2a^4}{\mathcal{H}_0^4} p_q &=
- \frac{\zeta_0^4\zeta^2}{96(1-\zeta)^2} + \frac{\zeta_0^4(\zeta-5\zeta^2)}{48(1-\zeta)} + \frac{\zeta_0^2\zeta^2}{6(2-\zeta-\zeta_0\zeta)^2} + \frac{(5\zeta_0^2\zeta^2 - 2\zeta(\zeta_0^2-\zeta_0))}{6(2-\zeta-\zeta_0\zeta)} 
 \nonumber \\
&+ \frac{(1-\zeta_0)(253+35\zeta_0)}{30240\zeta^2} + \frac{(1-\zeta_0)(109+35\zeta_0)}{15120\zeta}
\nonumber \\
& + \frac{1}{241920} \Big{[} -56227 + 18148\zeta_0 + 13386 \zeta_0^2  -12260\zeta_0^3 - 2737\zeta_0^4 \Big{]} 
\nonumber \\
& + \frac{\zeta}{60480} \Big{[} 867 + 12236\zeta_0 + 10246 \zeta_0^2 + 2404 \zeta_0^3 - 1183 \zeta_0^4 \Big{]} 
\nonumber \\
& + \frac{\zeta^2}{483840} \Big{[} -7183 + 46362\zeta_0 - 148073 \zeta_0^2 + 113180 \zeta_0^3 - 4745 \zeta_0^4 + 8586 \zeta_0^5 - 567 \zeta_0^6 \Big{]} 
\nonumber \\
& + \frac{3\zeta^3}{8960} \Big{[} -105 + 310\zeta_0 - 391 \zeta_0^2 + 84 \zeta_0^3 + 1041\zeta_0^4 - 106\zeta_0^5 + 7\zeta_0^6 \Big{]}
 \\
& + \frac{\zeta^4}{53760} \Big{[} 135 \!+\! 1890\zeta_0 \!-\! 1845\zeta_0^2 \!-\! 13500\zeta_0^3 \!+\! 5(3024\gamma_E\!-\!3635)\zeta_0^4\!+\! 954\zeta_0^5\!-\! 63\zeta_0^6 \Big{]}
\nonumber \\
& + \frac{9\zeta_0^4\zeta^4}{32} \ln \left[ 2a \left( \frac{\mu_f}{\mathcal{H}_0} \right) \frac{1-\zeta_0}{\zeta_0} \right]
+ \left( \frac{1}{6} + \frac{\zeta_0^4}{96} \right) \ln \left[ \frac{1-\zeta_0}{\tau\mathcal{H}_1} \right]
\nonumber \\
& + \Bigg{\{} \frac{1}{108\zeta^4} - \frac{4}{135\zeta} + \frac{1}{6} + \frac{\zeta_0^4}{96} - \frac{8\zeta^2}{27} + \zeta^4\left[ \frac{3}{20} + \frac{9\zeta_0^4}{32} \right] \Bigg{\}} \ln \left[ \frac{2(1-\zeta)}{(1-\zeta_0)\zeta}  \right] 
\nonumber \\
&+ \Bigg{\{}\!-\! \frac{1}{216\zeta^4} \!-\! \frac{1}{105\zeta^3} \!+\! \frac{-11 \!+\!\zeta_0^3}{540\zeta}  \!+\! \frac{5+4\zeta_0^3\!-\!\zeta_0^4}{192} \!+\! \zeta^2\left[\frac{697}{4320} \!-\! \frac{11\zeta_0^3}{216}  \!+\! \frac{3\zeta_0^4}{32} \!-\!\frac{\zeta_0^5}{40}  \!+\! \frac{\zeta_0^6}{432} \right]
\nonumber \\
& + \zeta^4 \left[ -\frac{4449}{71680} \!-\! \frac{141\zeta_0^3}{1280}  \!+\!\frac{45 \zeta_0^4}{1024}  \!+\!\frac{33\zeta_0^5}{640} \!-\!\frac{9\zeta_0^6}{256}  \!+\! \frac{9\zeta_0^7}{1792} \!-\!\frac{3\zeta_0^8}{10240} \right] \Bigg{\}} \ln \left[ 1  \!+\! \frac{2(1\!-\!\zeta)}{(1\!-\!\zeta_0)\zeta} \right] 
\nonumber \\
& + \Bigg{\{} - \frac{1}{216\zeta^4}  \!+\! \frac{1}{105\zeta^3} \!+\! \frac{27\!-\!\zeta_0^3}{540\zeta}  \!+\!\frac{-27+4\zeta_0^3\!-\!\zeta_0^4}{192} \!+\! \zeta^2 \left[ \frac{27}{160} \!-\!\frac{\zeta_0^3}{8}  \!+\!\frac{3\zeta_0^4}{32} \!-\!\frac{\zeta_0^5}{40}  \!+\! \frac{\zeta_0^6}{432} \right] 
\nonumber \\
&+ \zeta^4 \left[ -\frac{6561}{71680}  \!+\! \frac{243\zeta_0^3}{1280} \!-\!\frac{243\zeta_0^4}{1024}  \!+\! \frac{81\zeta_0^5}{640}\!-\! \frac{9\zeta_0^6}{256}  \!+\! \frac{9\zeta_0^7}{1792} \!-\!\frac{3\zeta_0^8}{10240} \right] \Bigg{\}} \ln\left[ \left| 1 \!-\!\frac{2(1\!-\!\zeta)}{(1\!-\!\zeta_0)\zeta} \right| 
\right]
\ .
\nonumber
\end{align}

\section*{Appendix E - Discussion of tensor fluctuations}

In this Appendix we present quantization of the graviton field on FLRW spaces 
and show one can relate its one-loop expectation value of the
energy-momentum tensor to the corresponding quantity for the 
minimally coupled massless scalar field.  

 The relevant graviton action can be obtained by simply perturbing
the Hilbert-Einstein action plus the matter action 
in perturbations of the metric to quadratic order in perturbations. 
For our purposes it is convenient to perturb the metric tensor $g_{\mu\nu}$ as
\begin{equation}
   g_{\mu\nu}(x) = a^2(\eta)\eta_{\mu\nu} + \delta g_{\mu\nu}(x)
\,,\qquad  \delta g_{\mu 0}(x) = 0
\,,\qquad  \delta g_{ij}(x) = a^2(\eta)\kappa h_{ij}(x)
\,,\qquad \kappa^2 = 16\pi G_N
\label{graviton perturbation}
\end{equation}
where $h_{ij}$ is traceless and transverse, 
\begin{equation}
\delta_{ij}h_{ij}= 0 =\partial_i h_{ij}
\ .
 \label{refhij}
\end{equation}
This {\it tensor} metric perturbation is invariant under linear 
coordinate transformations $x^\mu\rightarrow x^\mu + \xi^\mu(x)$ 
(also known as gravitational gauge transformations) and therefore constitutes
the physical graviton field, and in $D$ space-time dimensions 
has $D(D-3)/2$ components which are the physical polarizations of the 
graviton fields (in $D=4$ there are two polarizations, known as 
the $+$ and $\times$ polarizations). 

 The quadratic action for $h_{ij}$ in FLRW spaces is well known 
(see {\it e.g.}~\cite{ProkopecRigopoulos}),
\begin{align} \label{wavesaction}
S_{g}= - \frac{1}{4}\sum_{ij} \int d^D x 
\sqrt{-g} \, g^{\mu\nu}\partial_{\mu}h_{ij}\partial_{\nu}h_{ij} 
\,,\qquad g_{\mu\nu} = a^2(\eta) \eta_{\mu\nu}
\,.
\end{align}
This action is almost identical to the action of a minimally coupled 
massless scalar field. The difference is that $h_{ij}$ is not a scalar, 
but a spatial tensor with $\frac{1}{2}D(D-3)$ independent components.

 The canonical quantization of the graviton field on a FLRW background
proceeds by first calculating the canonically conjugate momenta,
\begin{align}
\pi^{ij}= \frac{\delta S_g}{\delta h'_{ij}} 
=  \frac{1}{2}a^{D-2} h_{ij}^{\prime},
\end{align}
and then promoting the fields into operators and
imposing canonical commutation relations
that makes use of the Dirac bracket
rather than the usual Poisson bracket~\cite{Dirac}. 
When this procedure for canonical quantization of constrained systems 
is applied to the graviton, the result is the following 
quantization rule,  
\begin{align}
& [\hat{h}_{ij}(\eta,\boldsymbol{x}), \hat{\pi}^{kl}(\eta,\boldsymbol{x}')]=
\frac{i}{2} 
\left(
P_{ik}P_{jl}+P_{il} P_{jk}-\frac{2}{D-2}P_{ij}P_{kl}
\right)\delta^{D-1}(\boldsymbol{x}-\boldsymbol{x}') 
\label{comm h} \ , \\
& [\hat{h}_{ij}(\eta,\boldsymbol{x}) , \hat{h}_{kl}(\eta,\boldsymbol{x}')] = 0 = [\hat{\pi}^{ij}(\eta,\boldsymbol{x}) , \hat{\pi}^{kl}(\eta,\boldsymbol{x}')]
\end{align}
where $P_{ij}=\delta_{ij}-\partial_i \partial_j  / \nabla^2$ is 
a transverse projector.
The purpose of the projector operator on the right hand side
of the quantization rule~(\ref{comm h}) is to ensure 
the tracelessness and transversality~(\ref{refhij}) of both 
$\hat{h}_{ij}$ and $\hat{\pi}^{ij}$.
This way only the $D(D-3)/2$ physical graviton polarization are quantized.
These commutation relations, together with the Heisenberg equation, lead
to the equation of motion for the graviton field operator,
\begin{equation}
\hat{h}_{ij}^{\prime\prime} + (D-2) \mathcal{H}\hat{h}_{ij}^{\prime}
- \boldsymbol{\nabla}^2 \hat{h}_{ij} = 0 \ .
\end{equation}
Furthermore, taking into account spatial isotropy of the quantum state  
we can expand the field operator in Fourier modes as follows,
\begin{align}
\hat{h}_{ij}(x)= \frac{a^{\frac{2-D}{2}}}{\sqrt{2}}
 \int \frac{d^{D-1} k}{(2\pi)^{D-1}} \sum_{\sigma} 
\bigg{\{} {\rm e}^{i\boldsymbol{k}\cdot \boldsymbol{x}} 
   \epsilon^{\sigma}_{ij}(\boldsymbol{k})
h(k,\eta) \hat{b}_{\sigma}(\boldsymbol{k})
 + {\rm e}^{-i\boldsymbol{k}\cdot \boldsymbol{x}} 
  \epsilon^{\sigma*}_{ij}(\boldsymbol{k})
     h^*(k,\eta) \hat{b}_{\sigma}^{\dag}(\boldsymbol{k})
\bigg{\}}  ,
\end{align}
where the sum over $\sigma$ goes over $\frac{1}{2}D(D-3)$ 
internal degrees of freedom (graviton polarizations). 
The annihilation operators, 
$\hat{b}(\sigma,\boldsymbol{k})$, and the creation operators, 
$\hat{b}^{\dag}(\sigma,\boldsymbol{k})$, satisfy
\begin{align} \label{comm b}
& [ \hat{b}_{\sigma}(\boldsymbol{k}) , 
\hat{b}^{\dag}_{\sigma'}(\boldsymbol{k}') ] 
= (2\pi)^{D-1} \delta_{\sigma \sigma'}  
\delta^{D-1}(\boldsymbol{k}-\boldsymbol{k}') 
\  \ , \ \ \  
[\hat{b}_{\sigma}(\boldsymbol{k}) , \hat{b}_{\sigma'}(\boldsymbol{k}') ]
 = [ \hat{b}^{\dag}_{\sigma}(\boldsymbol{k}) 
, \hat{b}^{\dag}_{\sigma'}(\boldsymbol{k}') ] = 0 \ , 
\end{align}
and the polarization tensor, $\epsilon_{ij}^{\sigma}(\boldsymbol{k})$, 
satisfies
\begin{align}
& \sum_{ij} \epsilon_{ij}^\sigma(\boldsymbol{k}) \epsilon_{ij}^{\sigma'}(\boldsymbol{k}) = \delta_{\sigma\sigma'} \ , \\
& \sum_{\sigma} \epsilon_{ij}^{\sigma}(\boldsymbol{k}) \epsilon_{kl}^{\sigma}(\boldsymbol{k}) = \frac{1}{2} \left[ \widetilde{P}_{ik} \widetilde{P}_{jl} + \widetilde{P}_{il} \widetilde{P}_{jk} - \frac{2}{D-2} \widetilde{P}_{ij} \widetilde{P}_{kl} \right] \ , \label{2nd sum}
\end{align}
where $\widetilde{P}_{ij} = \delta_{ij} - k_ik_j/k^2$ is the transverse projector in momentum space. The mode function, $h(k,\eta)$, satisfies 
the equation of motion
\begin{equation}
h''(k,\eta) + \left[ k^2 + f(\eta) \right]h(k,\eta) = 0 \ ,
\end{equation}
where, as in the case of the massless scalar field in~(\ref{f}),
$f(\eta) = -\frac{1}{4}(D-2) \left[ 2\mathcal{H}' 
+ (D-2)\mathcal{H}^2 \right]$. 
The properties \eqref{comm b}-\eqref{2nd sum} 
fix the Wronskian normalization of the graviton mode function to be 
the canonical one, 
\begin{equation}
W[h(k,\eta),h^*(k,\eta)] 
= h(k,\eta) h'^*(k,\eta) - h'(k,\eta) h^*(k,\eta) =  i
\ .
\end{equation}
The global Bunch-Davies vacuum state is defined by
\begin{align}
\hat{b}_{\sigma}(\boldsymbol{k}) | \Omega \rangle =0
\ \ \ (\forall \boldsymbol{k},\sigma)
\,,
\end{align}
and the entire Fock space of states is constructed by acting
with creation operators on this vacuum state.
From these results we see that the procedure for the choice of 
the graviton mode function is the same as the one outlined in the end
of section~\ref{Scalar field on FLRW}, where 
we defined the global Bunch-Davies state on a FLRW background.

 The energy-momentum tensor operator of the graviton field is
\begin{equation}
\hat{T}_{\mu\nu}(x) 
= \left. \frac{-2}{\sqrt{-g}} \frac{\delta S_g}{\delta g^{\mu\nu}(x)}
 \right|_{h_{ij}=\hat{h}_{ij}} 
= \frac12\sum_{i,j} \left[ \partial_{\mu} \hat{h}_{ij} 
   \partial_\nu \hat{h}_{ij} 
- \frac{1}{2}g_{\mu\nu} g^{\alpha\beta}  
  \partial_{\alpha} \hat{h}_{ij} \partial_\beta \hat{h}_{ij} \right] \ .
\end{equation}
Its one-loop expectation value on a FLRW background is diagonal, 
and turns out to be
\begin{align}
 \langle\Omega | \hat{T}_{00} |\Omega\rangle
& = \frac{D(D-3)}{2} \frac{a^{2-D}}{(4\pi)^{\frac{D-1}{2}}
 \Gamma\left( \frac{D-1}{2} \right)} 
 \\
&\times \int_0^{\infty}dk\, k^{D-2} \bigg{\{} |h'|^2 
+ \left[ k^2 + \frac{(D-2)^2}{4}\mathcal{H}^2 \right] |h|^2 
- \frac{D-2}{2}\mathcal{H} \frac{\partial}{\partial\eta}|h|^2 \bigg{\}} 
\ ,
\nonumber \\
 \langle\Omega | \hat{T}_{ij} |\Omega\rangle 
&= \frac{D(D-3)}{2} \frac{\delta_{ij} 
a^{2-D}}{(4\pi)^{\frac{D-1}{2}}\Gamma\left( \frac{D-1}{2} \right)}
 \\
&\times \int_0^{\infty}dk\, k^{D-2} \bigg{\{} |h'|^2 
 + \left[ -\frac{D-3}{D-1}k^2
 + \frac{(D-2)^2}{4}\mathcal{H}^2 \right] |h|^2 
- \frac{D-2}{2}\mathcal{H} \frac{\partial}{\partial\eta}|h|^2 \bigg{\}} \ .
\nonumber
\end{align}
We now see that the expectation value of the one-loop energy-momentum
tensor for gravitons is a multiple of the expectation value for the scalar
in Eqs.~(\ref{intT00}--\ref{intTij}).
This multiplication factor is $\frac{1}{2}D(D-3)$, which precisely accounts for
the difference in the number of internal degrees of freedom (polarizations)
between the graviton and the scalar. 
Therefore, in four space-time dimensions, the one-loop energy-momentum tensor
for gravitons will be twice as large as the scalar one.
Of course, after regularization there will remain finite free 
contributions coming from the $R^2$ counterterm
that will be of the same type as those of the scalar field. 
Therefore, the finite contributions from all massless scalar fields and those 
of the graviton can be subsumed into one finite coupling constant
$\alpha_f$ in Eq.~(\ref{counterterm:R2}), 
that in principle can be determined by measurements.

The discussion presented in this Appendix lends support to the claim made
in section~\ref{Discussion and conclusions}, 
where it is stated that the graviton contributes to the one-loop
energy-momentum tensor as two real scalar fields, such 
that we need not consider it separately.


\begin{thebibliography}{99}


\bibitem{KolbMatarreseRiotto}
E.~W.~Kolb, S.~Matarrese, A.~Riotto,
"On cosmic acceleration without dark energy,"
New~J.~Phys. {\bf8}, 322 (2006)
[arXiv:astro-ph/0506534].

\bibitem{BarausseMatarreseRiotto}
E.~Barausse, S.~Matarrese, A.~Riotto,
"The Effect of Inhomogeneities on the Luminosity Distance-Redshift Relation: 
 is Dark Energy Necessary in a Perturbed Universe?,"
[arXiv:astro-ph/0501152].

\bibitem{HirataSeljak}
C.~M.~Hirata and U.~Seljak,
"Can superhorizon cosmological perturbations explain 
 the acceleration of the universe?,"
Phys.\ Rev.\ D {\bf 72}, 083501 (2005)
[arXiv:astro-ph/0503582].

\bibitem{Abramo:2001dc}
  L.~R.~Abramo and R.~P.~Woodard,
  ``No one loop back reaction in chaotic inflation,''
  Phys.\ Rev.\ D {\bf 65} (2002) 063515
  [astro-ph/0109272].

\bibitem{Abramo:2001db}
  L.~R.~Abramo and R.~P.~Woodard,
  ``A Scalar measure of the local expansion rate,''
  Phys.\ Rev.\ D {\bf 65} (2002) 043507
  [astro-ph/0109271].

\bibitem{Abramo:1998hj}
  L.~R.~W.~Abramo and R.~P.~Woodard,
  ``One loop back reaction on power law inflation,''
  Phys.\ Rev.\ D {\bf 60} (1999) 044011
  [astro-ph/9811431].

\bibitem{Abramo:1998hi}
  L.~R.~W.~Abramo and R.~P.~Woodard,
  ``One loop back reaction on chaotic inflation,''
  Phys.\ Rev.\ D {\bf 60} (1999) 044010
  [astro-ph/9811430].

\bibitem{Geshnizjani:2002wp}
  G.~Geshnizjani and R.~Brandenberger,
  ``Back reaction and local cosmological expansion rate,''
  Phys.\ Rev.\ D {\bf 66} (2002) 123507
  [gr-qc/0204074].

\bibitem{Geshnizjani:2003cn}
  G.~Geshnizjani and R.~Brandenberger,
  ``Back reaction of perturbations in two scalar field inflationary models,''
  JCAP {\bf 0504} (2005) 006
  [hep-th/0310265].

\bibitem{Abramo:2001dd}
  L.~R.~Abramo and R.~P.~Woodard,
  ``Back reaction is for real,''
  Phys.\ Rev.\ D {\bf 65} (2002) 063516
  [astro-ph/0109273].

\bibitem{Janssen:2009nz}
  T.~M.~Janssen and T.~Prokopec,
  ``Regulating the infrared by mode matching: 
    A Massless scalar in expanding spaces with constant deceleration,''
  Phys.\ Rev.\ D {\bf 83} (2011) 084035
  [arXiv:0906.0666 [gr-qc]].

\bibitem{Janssen:2008dw}
  T.~Janssen and T.~Prokopec,
  ``The Graviton one-loop effective action in cosmological space-times 
  with constant deceleration,''
  Annals Phys.\  {\bf 325} (2010) 948
  [arXiv:0807.0447 [gr-qc]].



\bibitem{Koivisto:2010pj}
  T.~S.~Koivisto and T.~Prokopec,
  ``Quantum backreaction in evolving FLRW spacetimes,''
  Phys.\ Rev.\ D {\bf 83} (2011) 044015
  [arXiv:1009.5510 [gr-qc]].

\bibitem{Prokopec:1996rr}
  T.~Prokopec and T.~G.~Roos,
  ``Lattice study of classical inflaton decay,''
  Phys.\ Rev.\ D {\bf 55} (1997) 3768
  [hep-ph/9610400].

\bibitem{Capper:1974ic}
  D.~M.~Capper and M.~J.~Duff,
  ``Trace anomalies in dimensional regularization,''
  Nuovo Cim.\ A {\bf 23} (1974) 173.

\bibitem{Duff:1977ay}
  M.~J.~Duff,
  ``Observations on Conformal Anomalies,''
  Nucl.\ Phys.\ B {\bf 125} (1977) 334.

\bibitem{Birrell:1982ix}
  N.~D.~Birrell and P.~C.~W.~Davies,
  ``Quantum Fields In Curved Space,''
  Cambridge, Uk: Univ. Pr. (1982) 340p

\bibitem{Parker1969}
 L.~Parker, 
 ``Quantized Fields and Particle Creation in Expanding Universes. I,''
 Phys.~Rev. \textbf{183},  1057 (1969)

\bibitem{FP}
  L.~H.~Ford and L.~Parker,
  ``Infrared Divergences In A Class Of Robertson-Walker Universes,''
  Phys.\ Rev.\  D {\bf 16} (1977) 245.

\bibitem{ProkopecRigopoulos}
T.~Prokopec and G.~Rigopoulos,
Phys. Rev. D \textbf{82}, 023529  (2010)
[arXiv:1004.0882].

\bibitem{Dirac}
P.~A.~M.~Dirac, 
``Lectures on Quantum Mechanics,''
New York, USA: Belfer Graduate School of Science (1964) 151p



\end{thebibliography}
\end{document}